\documentclass{article}

\PassOptionsToPackage{numbers, compress}{natbib}

\usepackage[final]{neurips_2022}

\usepackage[utf8]{inputenc}    
\usepackage[T1]{fontenc}         
\usepackage{url}                           
\usepackage{booktabs}             
\usepackage{amsfonts}             
\usepackage{nicefrac}                
\usepackage{microtype}           
\usepackage{xcolor}                   
\definecolor{lightpurple}{HTML}{9D3DCF}
\definecolor{darkred}{HTML}{E85642}
\definecolor{indigoblue}{HTML}{1280B0}
\usepackage[colorlinks=true, urlcolor=lightpurple, linkcolor=darkred, citecolor=indigoblue]{hyperref}

\usepackage{amsmath}
\usepackage{amssymb}
\usepackage{array}
\usepackage{threeparttable}
\usepackage{bbm}
\usepackage{bm}
\usepackage{caption}
\usepackage{color}
\usepackage{epstopdf}
\usepackage{graphicx}
\usepackage{multirow}
\usepackage{stfloats}
\usepackage{subfigure}
\usepackage{textcomp}
\usepackage{wrapfig}
\usepackage[all]{xy}

\usepackage{algorithmic}
\usepackage{algorithm}

\newcommand{\Omegav}[0]{\ensuremath{\boldsymbol{\Omega}} }

\newcommand{\alphav}[0]{\ensuremath{\boldsymbol{\alpha}} }

\newcommand{\cdotv}[0]{\ensuremath{\boldsymbol{\cdot}}}

\title{HyperMiner: Topic Taxonomy Mining with Hyperbolic Embedding} 

\author{%
  Yishi Xu, Dongsheng Wang, Bo Chen\thanks{Corresponding author} , Ruiying Lu, Zhibin Duan \\
  National Laboratory of Radar Signal Processing, Xidian University, Xi’an, China \\
  \texttt{xuyishi@stu.xidian.edu.cn, bchen@mail.xidian.edu.cn} \\
  \And 
  Mingyuan Zhou \\
  McCombs School of Business, The University of Texas at Austin, USA \\
  \texttt{mingyuan.zhou@mccombs.utexas.edu} \\
}

\begin{document}

\maketitle

\begin{abstract}
Embedded  topic models are able to learn interpretable topics even with large and heavy-tailed vocabularies. 
However, they generally hold the Euclidean embedding space assumption, leading to a basic limitation in capturing hierarchical relations. 
To this end, we present a novel framework that introduces hyperbolic embeddings to represent words and topics. 
With the tree-likeness property of hyperbolic space, 
the underlying semantic hierarchy among words and topics can be better exploited to mine more interpretable topics.
Furthermore, due to the superiority of hyperbolic geometry in representing hierarchical data, 
tree-structure knowledge can also be naturally injected to guide the learning of a topic hierarchy.
Therefore, we further develop a regularization term based on the idea of contrastive learning to inject prior structural knowledge efficiently. 
Experiments on both topic taxonomy discovery and document representation demonstrate that 
the proposed framework achieves improved performance against existing embedded topic models. 
\end{abstract}

\section{Introduction}
\label{introduction}
With a long track record of success in a variety of applications \cite{wang2007topical, mimno2009polylingual, rubin2012statistical, jiang2015author, wang2018topic, jelodar2020deep}, topic models have emerged as one of the most powerful tools for automatic text analysis. Typically, given a collection of documents, a topic model aims to identify a group of salient topics by capturing common word co-occurrence patterns. Despite their popularity, traditional topic models such as Latent Dirichlet Allocation (LDA) \cite{blei2003latent} and its variants \cite{griffiths2003hierarchical, blei2006correlated, mcauliffe2007supervised, zhou2012beta, paisley2014nested} are plagued by complicated posterior inference, presenting a challenge to create deeper and more expressive models of text. Fortunately, recent developments of Variational AutoEncoders (VAEs) and Autoencoding Variational Inference (AVI) \cite{kingma2013auto, rezende2014stochastic} have shed light on this problem, resulting in the proposal of a series of Neural Topic Models (NTMs) \cite{miao2016neural, srivastava2017autoencoding, zhang2018whai, nan2019topic}. With better flexibility and scalability, NTMs have gained increasing research interest over the past few years. 

\begin{figure}[tb]
\centering
\setlength{\belowcaptionskip}{-0.6cm}
\includegraphics[width=0.98\textwidth]{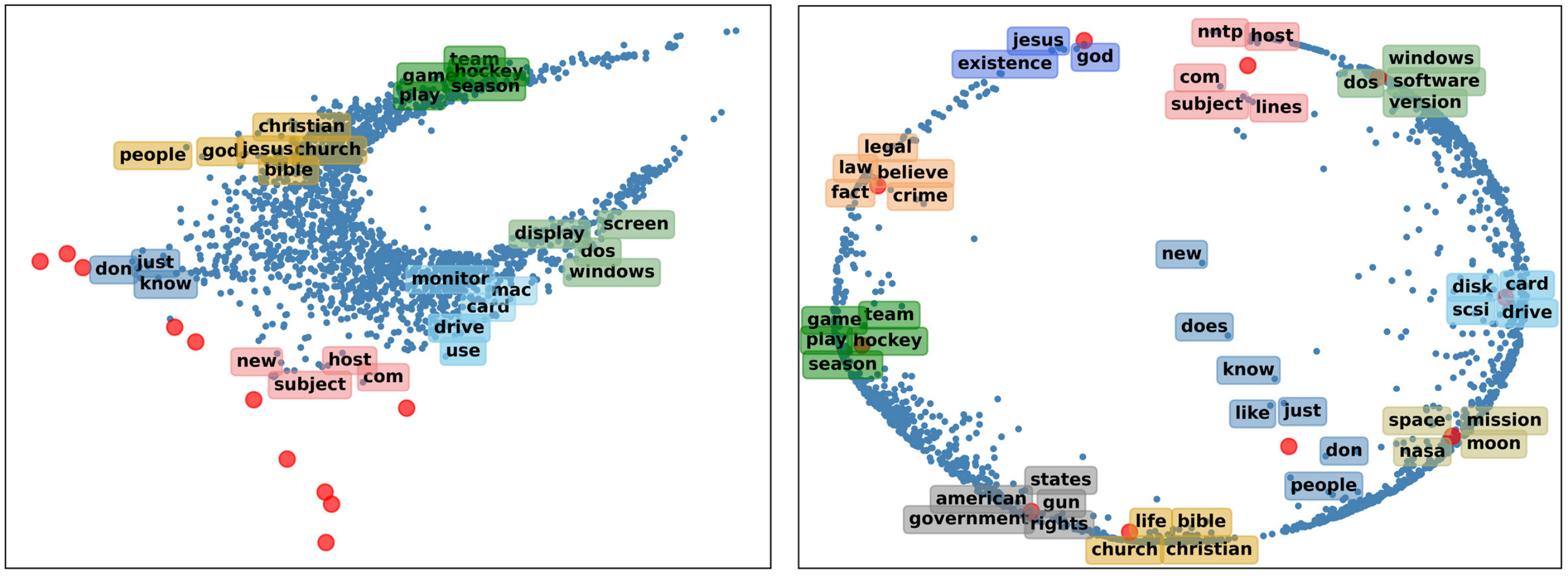}
\captionsetup{font={small}}
\caption{Visualization of 2D Euclidean embedding space (left) and 2D hyperbolic embedding space (right) learned by ETM. 
Red points denote topic embeddings, blue points represent word embeddings.}
\label{fig:2d_embedding_space}
\end{figure}

Parallel to neural topic modeling, the idea of bringing word embeddings \cite{mikolov2013efficient, mikolov2013distributed} 
into topic models has also attracted much attention. 
Considering the large performance degradation over short texts due to limited word co-occurrence information, 
some early works \cite{petterson2010word, nguyen2015improving, li2016topic} exploit word embeddings 
as complementary metadata and incorporate them into the generative process of topic models. 
Recently, more flexible ways \cite{zhao2018inter, wete} of combining word embeddings have been explored thanks to the development of NTMs.
For example, \citet{bianchi2020pre} use word embeddings directly as part of the encoder's input.
In particular, a novel one called Embedded Topic Model (ETM) \cite{dieng2020topic} stands out 
for its  performance as well as the elegant way it integrates word embeddings.
Specifically, by representing topics as points in the word embedding space, 
ETM assigns probabilities to words based on their (inner product) distances from each topic embedding. 
As a result, semantically related words tend to fall around the same topic, 
thus facilitating the discovery of more interpretable topics.

Under the inspiration of ETM, \citet{duan2021sawtooth} have extended a similar idea to 
hierarchical topic modeling and proposed SawETM.  
In addition to mapping words and hierarchical topics into a shared embedding space, 
SawETM has also developed a unique Sawtooth Connection module to capture the dependencies 
between the topics at different layers, which, on the other side, empowers it to support a deep network structure. 
While achieving promising results, both ETM and SawETM hold the Euclidean embedding space assumption, 
leading to a fundamental limitation that their ability to model complex patterns (akin to social networks, 
knowledge graphs, and taxonomies) is inherently bounded by the dimensionality of the embedding 
space \cite{nickel2017poincare, nickel2014reducing}. 
As a consequence, the underlying semantic hierarchy among the words and topics
can hardly be expressed adequately in a relatively low-dimensional embedding space, 
as illustrated on the left side of Figure~\ref{fig:2d_embedding_space}.

Apart from the difficulty in capturing the implicit semantic hierarchy, 
another concomitant problem is the dilemma of incorporating explicit structural knowledge.
Assuming we have a prior taxonomy of concepts and wish to use it to guide the learning of hierarchical topics,
it is challenging to preserve the structure between concepts in Euclidean space by constraining the word and topic embeddings.
To cope with this issue, TopicNet \cite{duan2021topicnet} employs the Gaussian-distributed embeddings
as a substitute for the vector embeddings to represent words and topics.
As such, the prior knowledge of hypernym relations between concepts 
could be naturally injected via the encapsulation of probability densities.
However, maintaining the semantic hierarchy in such an embedding space still suffers from a certain degree of distortion, 
as it poses a 
challenge
to the optimization of KL divergence between distributions.
Furthermore, the introduction of Gaussian-distributed embeddings entails a great demand on memory, 
limiting its potential scalability to large vocabularies and high-dimensional embedding spaces.

To overcome the above shortcomings brought by Euclidean embedding space,
we propose to compute embeddings in hyperbolic space. 
Distinguished by the tree-likeness properties \cite{gromov1987hyperbolic, hamann2018tree}, 
hyperbolic space has been consistently shown to be superior in modeling hierarchical data compared to Euclidean space 
\cite{ganea2018hyperbolic, sala2018representation, tifrea2018poincar, cho2019large}. 
By measuring the distance between words and topics in hyperbolic embedding space, 
the model is encouraged to better capture the underlying semantic hierarchy among words. 
As shown on the right side of Figure~\ref{fig:2d_embedding_space}, 
some general words such as ``new'' and ``does'' fall around the center, 
they stay close to all other points because they often co-occur with other words. 
While more specific words like ``moon'' and ``nasa'' fall near the boundary and are only close to the nearby points.
Moreover, hyperbolic space also provides a better platform to inject prior structural knowledge, 
since hierarchical relations can be effectively preserved by imposing constraints on the distance between word and topic embeddings. 
In a nutshell, the main contributions of this paper are as follows:
\begin{itemize}
\item We propose to compute the distance between topics and words in hyperbolic embedding space 
on the basis of existing embedded topic models, which is beneficial to both the mining of implicit semantic hierarchy and the incorporation of explicit structural knowledge.
\item We design a node-level graph representation learning scheme that can inject prior structural knowledge to effectively guide the learning of a meaningful topic taxonomy.
\item Extensive experiments on topic quality and document representation demonstrate that the proposed approach achieves competitive performance against baseline methods.
\end{itemize}

\section{Background}
\label{background}

\subsection{Embedded topic model}
ETM \cite{dieng2020topic} is a neural topic model that builds on two main techniques: LDA \cite{blei2003latent} and word embeddings \cite{mikolov2013efficient, mikolov2013distributed}. To marry the probabilistic topic modeling of LDA with the contextual information 
brought by word embeddings, ETM maintains vector representations of both words and topics and uses them to derive the per-topic distribution over the vocabulary.
Specifically, consider a corpus with $V$ distinct terms comprising the vocabulary, 
we denote the word embedding matrix as $\bm{\rho} \in \mathbb{R}^{D \times V}$, 
where $D$ is the dimensionality of the embedding space. 
For each topic, there is also an embedding representation $\bm{\alpha}_k \in \mathbb{R}^D$, 
then ETM defines the per-topic distribution $\bm{\beta}_k \in \mathbb{R}^V$ over the vocabulary as
\begin{equation}
    \begin{split}
        \bm{\beta}_k = \mbox{Softmax}\left(\bm{\rho}^\top\bm{\alpha}_k\right)
    \end{split}
\end{equation}
With the above definition, ETM specifies a generative process analogous to LDA.
Let $\bm{w}_{jn} \in \{1 , ... , V \}$ denote the $n^{th}$ word in the $j^{th}$ document, 
the generative process is as follows.
\begin{enumerate}
    \item Draw topic proportions $\bm{\theta}_j \sim \mathcal{LN}\left(\bm{0}, \bm{I}\right)$. \vspace{-0.05cm}
    \item For each word $n$ in the document: \vspace{-0.06cm}
    \begin{enumerate}
        \item Draw topic assignment $\bm{z}_{jn} \sim \mbox{Cat}\left(\bm{\theta}_j\right)$. 
        \item Draw word $\bm{w}_{jn} \sim \mbox{Cat}\left(\bm{\beta}_{\bm{z}_{jn}}\right)$.
    \end{enumerate}
\end{enumerate}
Where $\mathcal{LN}(\cdot)$ in step 1 denotes the logistic-normal distribution \cite{atchison1980logistic},
which transforms a standard Gaussian random variable to the simplex.
By taking the inner product of the word embedding matrix $\bm{\rho}$ and the topic embedding $\bm{\alpha}_k$ to derive $\bm{\beta}_k$, 
the intuition behind ETM is that semantically related words will be assigned to similar topics. 
With this property, ETM has been demonstrated to improve the quality of the learned topics, especially in the presence of large vocabularies. 
Like most NTMs, ETM is fitted via an efficient amortized variational inference algorithm.

\subsection{Hyperbolic geometry}
In this part, we briefly review some key concepts on hyperbolic geometry. 
A comprehensive and in-depth description can be found in \citet{lee2013smooth} and \citet{ nickel2018learning}.
Under the mathematical framework of Riemannian geometry, 
hyperbolic geometry specializes in the case of constant negative curvatures. 
Intuitively, the hyperbolic space can be understood as a continuous version of trees:
the volume of a ball expands exponentially with its radius,
just as how the number of nodes in a binary tree grows exponentially with its depth.
Mathematically, there exist multiple equivalent models for hyperbolic space with different definitions and metrics. 
Here, we consider two representative ones in light of optimization simplicity and stability: 
Poincaré ball model \cite{nickel2017poincare} and the Lorentz model \cite{nickel2018learning}.
\vspace{-0.1cm}
\paragraph{Poincaré ball model}
The Poincaré ball model of an $n$-dimensional hyperbolic space with curvature $C$ ($C<0$) is defined by the Riemannian manifold 
$\mathcal{P}^n = \left(\mathcal{B}^n, g_p\right)$, where $\mathcal{B}^n = \{\bm{x} \in \mathbb{R}^n: {\Vert \bm{x} \Vert} < {1/\sqrt{\lvert C \rvert}}\}$ 
is the open $n$-dimensional ball with radius $1/\sqrt{\lvert C \rvert}$ and $g_p$ is the metric tensor that can be converted from
the Euclidean metric tensor $g_e=I$ as
\begin{equation}
    \begin{split}
        g_p\left( \bm{x} \right) = \left(\frac{2}{1 + C{\Vert \bm{x} \Vert}^2} \right)^2{g_e}
    \end{split}
\end{equation}
\paragraph{Lorentz model}
The Lorentz model (also named Hyperboloid model) of an $n$-dimensional hyperbolic space with curvature $C$ ($C<0$) is 
defined by the Riemannian manifold $\mathcal{L}^n = (\mathcal{H}^n, g_l)$,
where $\mathcal{H}^n = \{\bm{x} \in \mathbb{R}^{n+1}: {\langle \bm{x}, \bm{x} \rangle}_\mathcal{L}=1/C\}$ 
and $g_l=\mbox{diag}\left( [-1,\bm{1}^\top_n] \right)$. 
${\langle \cdotv, \cdotv \rangle}_\mathcal{L}$ denote the \textit{Lorentzian inner product}.
Let $\bm{x}, \bm{y} \in \mathbb{R}^{n + 1}$, the Lorentz inner product induced by $g_l$ is calculated as
\begin{equation}
    \begin{split}
        {\langle \bm{x}, \bm{y} \rangle}_\mathcal{L} = {\bm{x}^\top}{g_l}{\bm{y}} = - {x_0}{y_0} + \sum_{i=1}^n{x_i}{y_i}
    \end{split}
\end{equation}
An intuitive illustration of the equivalence between the Poincaré ball model and the Lorentz model and some other related operations in hyperbolic space will be introduced in Appendix~\ref{appendC}.

\section{Topic Taxonomy Mining with Hyperbolic Embedding}
\label{sec3}
In this section, we elaborate on how the introduced hyperbolic embeddings facilitate the mining of implicit 
semantic hierarchy and the incorporation of explicit tree-structure knowledge, both of which encourage
the model to find more interpretable topics\footnote{Our code is available at \url{https://github.com/NoviceStone/HyperMiner}}.

\subsection{Hierarchical topic modeling in hyperbolic space}
\label{sec31}
Theoretically, the idea of representing words and topics in hyperbolic space is orthogonal to 
a wide range of topic models employing the word embeddings technique. 
To provide the foundation for the subsequent injection of structural knowledge,
we here apply our method to a hierarchical embedded topic model SawETM \cite{duan2021sawtooth}.
SawETM utilizes the adapted Poisson gamma belief network (PGBN) \cite{zhou2016augmentable}
as its generative module (decoder) and decomposes the topic matrices into 
the inner product of topic embeddings at adjacent layers. 
The novelty of our method lies in that the hierarchical relations
can be better reflected by the distances between embeddings in hyperbolic space.
Mathematically, the generative model with $L$ latent layers is formulated as \vspace{1mm}
\begin{equation} \label{generative_model} \small 
\begin{split}
& \bm{\theta}_j^{\left(L\right)} \sim \mbox{Gam}\left(\bm{\gamma}, e_j^{\left(L + 1\right)}\right), \bm{\theta}_j^{\left(l\right)} \sim 
\mbox{Gam}\left(\bm{\Phi}^{\left(l + 1\right)} \bm{\theta}_j^{\left(l + 1\right)}, e_j^{\left(l + 1\right)}\right), \ldots, \\
& \bm{\theta}_j^{\left(1\right)} \sim \mbox{Gam}\left(\bm{\Phi}^{\left(2\right)} \bm{\theta}_j^{\left(2\right)}, e_j^{(2)}\right), \bm{x}_j \sim \mbox{Pois}\left({\bm{\Phi}^{\left(1\right)}}\bm{\theta}_j^{ \left(1\right)}\right),  \\
& \bm{\Phi}^{\left(l\right)}=\mbox{Softmax}\left(\mathcal{S}\left(\bm{\alpha}^{\left(l-1\right)}, \bm{\alpha}^{\left(l\right)}\right)\right) \\
\end{split}
\end{equation}
The above formula clearly describes how the multi-layer document representation is generated
via a top-down process. Specifically, the latent representation of the top layer $\bm{\theta}_j^{(L)}$ 
is sampled from a fixed gamma prior distribution, then at each intermediate layer $l$  
the latent units $\bm{\theta}_j^{(l)} \in \mathbb{R}^{K_l}$ are factorized into 
the product of the factor loading matrix $\bm{\Phi}^{(l+1)} \in \mathbb{R}^{K_l \times K_{l+1}}$ 
and latent units $\bm{\theta}_j^{(l+1)} \in \mathbb{R}^{K_{l+1}}$ of the above layer.
Until the bottom layer, the observation of word count vector $\bm{x}_j \in \mathbb{Z}^V$
is modeled as the Poisson distribution.
Note that the subscript $j$ denotes the document index and some other variables
$\bm{\gamma}, e_j^{(L+1)}, \ldots, e_j^{(2)}$ are hyperparameters. 
Especially, the factor loading matrix $\bm{\Phi}^{(l)}$ of layer $l$ is derived based on 
the distance between the topic embeddings at two adjacent layers, i.e., 
$\bm{\alpha}^{(l-1)} \in \mathbb{R}^{K_{l-1} \times D}$ and $\bm{\alpha}^{(l)} \in \mathbb{R}^{K_l \times D}$.
Note that $\bm{\alpha}^{(0)} \in \mathbb{R}^{V \times D}$ represents the word embeddings.
Since all embeddings are projected into the hyperbolic space to fully explore 
the underlying semantic hierarchy among the words and topics,
we design our similarity score function as \vspace{1mm}
\begin{equation} \label{Hyp-Dist} \small
\begin{split}
& \mathcal{S}\left(\bm{x}, \bm{y}\right) = -d_\mathcal{P}\left(\bm{x}, \bm{y} \right) = 
\frac{-1}{\sqrt{\lvert C \rvert}} \mbox{arcosh}\left(1 - \frac{2C{\Vert \bm{x} - \bm{y} \Vert}^2}
{\left(1 + C{\Vert \bm{x} \Vert}^2\right)\left(1 + C{\Vert \bm{y} \Vert}^2\right)}\right)\\
& \mathcal{S}\left( \bm{x}, \bm{y} \right) = -d_\mathcal{L}(\bm{x}, \bm{y}) = 
\frac{-1}{\sqrt{\lvert C \rvert}}\mbox{arcosh}\left( C {\langle \bm{x}, \bm{y} \rangle}_\mathcal{L} \right)\\
\end{split}
\end{equation}
Where $d_\mathcal{P}(\cdotv, \cdotv)$ and $d_\mathcal{L}(\cdotv, \cdotv)$ are the distance functions
of the Poincaré ball model and the Lorentz model, respectively.
As the two models of hyperbolic space are mathematically equivalent,
we take the Poincaré ball as an example for analysis. 
Eq.~(\ref{Hyp-Dist}) shows that the distance changes
smoothly with respect to the norm of $\bm{x}$ and $\bm{y}$.
This locality plays a crucial role in learning continuous embeddings of hierarchies.
For instance, the origin of $\mathcal{B}^n$ has a zero norm, it would have relatively small distance to all other points, which exactly corresponds to the root node of a tree. On the other hand, those points close to the boundary of the ball have a norm close to one, so the distance between them grows quickly, which 
properly reflects the relationships between the leaf nodes of a tree.

\begin{figure}[tb]
\centering
\setlength{\belowcaptionskip}{-0.3cm}
\includegraphics[width=\textwidth]{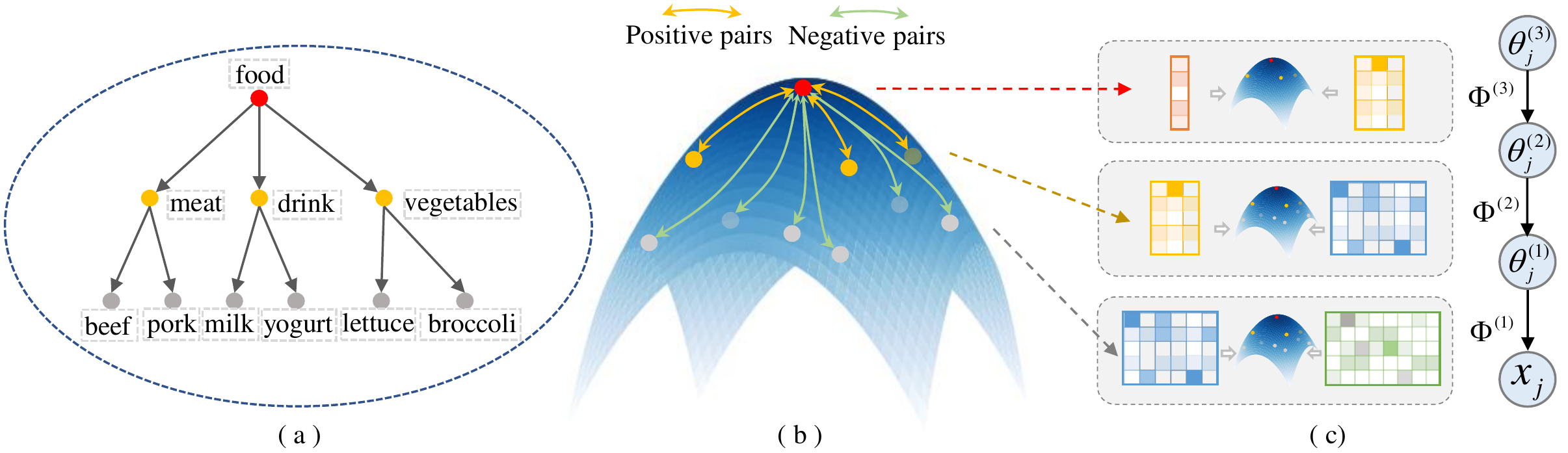}
\captionsetup{font={small}}
\caption{Overview of knowledge guided topic taxonomy discovery.
(a) The prior concept taxonomy constructed from vocabulary; 
(b) Illustration of the strategy for picking positive pairs and negative pairs in the hyperbolic embedding space;
(c) The hierarchical generative model whose factor loading matrices are derived based on the hyperbolic distances between topic embeddings at two adjacent layers.}
\label{fig:overview}
\end{figure}

\subsection{Knowledge guided topic taxonomy discovery}
Hierarchical structures are ubiquitous in knowledge representation and reasoning \cite{nickel2018learning}.
Particularly, mining a set of meaningful topics organized into a hierarchy from massive text corpora 
is intuitively appealing, as it allows users to easily access the information of their interest.
However, most existing hierarchical topic models struggle to realize this goal without any supervision,
and some appropriate guidance with prior structural knowledge proves to be helpful for mitigating this issue \cite{duan2021topicnet,wang2022knowledge}.

\paragraph{Form of prior knowledge}
We assume the prior knowledge takes the form of a concept taxonomy,
which is compatible with the deep structure of the proposed hierarchical topic model in Section~\ref{sec31}.
In detail, the taxonomy exhibits a top-down reification process 
and concepts between two adjacent layers with connections follow the hypernym relation, 
as shown in Figure~\ref{fig:overview}(a).
Meanwhile, to keep the taxonomy consistent with the corresponding dataset,
we construct it by traversing each word in the vocabulary to
find its ancestors along the hypernym paths provided by WordNet \cite{miller1995wordnet}.

\paragraph{Hyperbolic contrastive loss} \label{sec32}
Although the paradigm of contrastive learning has been successfully applied to graph representation learning 
in Euclidean space \cite{you2020graph, hassani2020contrastive, zhu2020deep, peng2020graph}, 
those developed contrastive algorithms are not directly applicable in our case.
On the one hand, we focus more on learning node representations while 
not destroying their prior hierarchical structures.
On the other hand, hyperbolic space possesses distinctive properties ($e.g.$, hierarchical awareness 
and spacious room) compared to its Euclidean counterpart.
Consequently, to accommodate the two differences, 
we design a node-level hyperbolic contrastive loss such that 
the prior knowledge can be effectively injected as an inductive bias to influence the learning of topic taxonomy. 

Specifically, we set the number of topics at each layer of the hierarchical topic model to be
the same as the number of concepts at each layer of the taxonomy,
then each topic is assigned a corresponding concept as its semantic prior. 
Since concepts are connected in the taxonomy, 
such relations can be transfered accordingly between topics, 
which provide the basis for picking positive and negative pairs among the topic and word embeddings.
Let $\mathcal{T} = \{\bm{\alpha}^{(l)}\}_{l=0}^L$ denote the set of all embeddings,
then each embedding $\bm{\alpha}_i^{(l)} \in \mathbb{R}^D$ is associated with 
two groups of embeddings as the positive samples and the negative samples, respectively.
Then the average hyperbolic contrastive loss is defined as (we omit the superscript $l$ for simplicity of notation)
\begin{equation} \label{Contra-Loss} \small
\begin{split}
  &  \mathcal{L}_{\rm Contra} = \mathbb{E}_{\bm{\alpha_i} \in \mathcal{T}} \left[ -\mbox{log}
  \frac{ \mbox{exp}\left( \mathcal{S}(\bm{\alpha}_i, \bm{\alpha}_i^+) / \tau\right)}
  {\mbox{exp}\left( \mathcal{S}(\bm{\alpha}_i, \bm{\alpha}_i^+) / \tau\right) + \sum_{\alphav_i^{-} \in \mathcal{Q}(\alphav_i)}
  \mbox{exp}\left( \mathcal{S}(\bm{\alpha}_i, \bm{\alpha}_i^{-}) / \tau \right)} \right]    \\
\end{split}
\end{equation}
where $\mathcal{S\left(\cdotv, \cdotv\right)}$ is the similarity score function defined in Eq.~(\ref{Hyp-Dist}) and $\tau$ is the temperature parameter. Note that $\alphav_i^+$ is a positive sample drawn from $\mathcal{P}(\alphav_i)$ and $\mathcal{Q}(\alphav_i)$ is the set of negative samples.

\paragraph{Sampling strategy }
Inspired by the homophily property ($i.e.$, similar actors tend to associate with each other) in many graph networks \cite{nickel2017poincare},
we take one-hop neighbors of each anchor, $i.e.$, its parent node and its child nodes as positive samples to maintain the hierarchical semantic information. For the negative samples, we select $m$ embeddings from the non-first-order neighbors that have the highest similarity scores with the anchor embedding.

\subsection{Training objective}
As most existing NTMs can be viewed as the extensions of the framework of VAEs \cite{kingma2013auto, rezende2014stochastic}, 
they generally develop a similar training objective to VAEs, which is to maximize the Evidence Lower BOund (ELBO).
For our generative model, the ELBO of each document can be derived as
\begin{equation} \label{ELBO}\small
\begin{split}
  &  \mathcal{L}_{\rm ELBO} =  - \sum_{l=1}^L D_{KL} \left[ q(\bm{\theta}_j^{(l)} \vert - ) \Vert p(\bm{\theta}_j^{(l)} \vert \bm{\Phi}^{(l+1)}, \bm{\theta}_j^{(l+1)}) \right] + \mathbb{E}_{q(\bm{\theta}_j^{(1)} \vert -)} \left[ \ln {p( \bm{x}_j \vert \bm{\Phi}^{(1)}, \bm{\theta}_j^{(1)} )} \right]\\
\end{split}
\end{equation}

\begin{algorithm}[tb] 
   \caption{Knowledge-Guided Topic Taxonomy Mining}
   \label{alg:example}
\begin{algorithmic}
   \STATE Input: mini-batch size $B$, number of layers $T$, adjacent matrix $A$ built from concept taxonomy. 
   \STATE Initialize the variational network parameters $\small \Omegav$ and the word and topic embeddings $\small \{ \bm{\alpha}^{(l)} \}_{l=0}^L$;
   \WHILE{not converged}
   \STATE1. Randomly draw a batch of samples $\small \left\{ \bm{x}_j \right\}_{j=1} ^{B}$;
   \STATE2. Infer variational posteriors for the latent variables of different layers $\small \{ \bm{\theta}_j^{(l)}  \}_{j=1, l=1}^{B, L};$ 
   \STATE3. Derive factor loading matrices $\small \left \{ \bm{\Phi}^{(l)}\right \}_{l=1}^{L}$ using $ \small \left \{ \bm{\alpha}^{(l)} \right \}_{l=0}^L$ based on Eq.~\eqref{generative_model};
   \STATE4. Compute the ELBO on the joint marginal likelihood of $\small  \left \{  \bm{x}_j \right\}_{j=1} ^{B}$ based on Eq.~\eqref{ELBO};  
   \STATE5. Compute the hyperbolic contrastive loss using $\small \left \{ \bm{\alpha}^{(l)} \right \}_{l=0}^L$ and $A$ according to Eq.~\eqref{Contra-Loss};
   \STATE6. Update  $\small \Omegav$ and $\small \left \{ \bm{\alpha}^{(l)} \right \}_{l=0}^L$ using gradients $\small \nabla {}_{\Omegav, \bm{\alpha}^{(l)}}\mathcal{L} \left({\Omegav, \{  \bm{\alpha}^{(l)}\}_{l=0}^L; \left\{ \bm{x}_j \right\}_{j=1} ^{B } } \right)$;
   \ENDWHILE
\end{algorithmic}
\end{algorithm} 
\setlength{\textfloatsep}{0.2cm}
where the first term is the Kullback--Leibler divergence that constrains the approximate posterior  $q(\bm{\theta}_j^{(l)} \vert - )$ to be close to the prior $p(\bm{\theta}_j^{(l)})$, and the second term denotes the expected log-likelihood or reconstruction error. Considering that our generative model employs the gamma-distributed latent variables, 
it brings the difficulty of reparameterizing a gamma-distributed random variable when we design a sampling-based inference network. Therefore, we instead utilize a Weibull distribution to approximate the conditional posterior inspired by \citet{zhang2018whai}, as the analytic KL expression and efficient reparameterization make it easy to estimate the gradient of ELBO with respect to network parameters. The implementation details of our variational encoder is described in Appendix~\ref{appendB}.

Furthermore, to inject the prior knowledge to guide the learning of a topic taxonomy, 
we train the ELBO jointly with a regularization term specified by the proposed contrastive loss in Section~\ref{sec32}
\begin{equation} \label{Total-Loss}
\begin{split}
  &  \mathcal{L} = {\mathcal{L}_{\rm ELBO}} + \lambda {\mathcal{L}_{\rm Contra}}    \\
\end{split}
\end{equation}
where $\lambda$ is the hyper-parameter used to control the impact of the regularization term, whose detailed effect is investigated in Appendix~\ref{appendD}.
We summarize our complete learning procedure in Algorithm~\ref{alg:example}.  
\vspace{-5mm}
\begin{table}[!h]
  \caption{Statistics of the datasets}
  \label{datasets}
  \setlength{\tabcolsep}{8pt} 
  \renewcommand{\arraystretch}{1.25} 
  \centering
  \small
  \begin{tabular}{ccccc}
    \toprule
    &Number of docs      &Vocabulary size     &Total number of words      &Categories \\
    \cmidrule(r){2-5}
    20NG     &18,846        &8,581         &1,446,490         & 20   \\
    TMN       &32,597        &13,368      &592,973             & 7 \\
    WIKI       &28,472        &20,000      &3,719,096         & N/A  \\
    RCV2      &804,414      &7,282        &60,209,009      & N/A \\
    \bottomrule
  \end{tabular}
\end{table}
\vspace{-5mm}
\section{Experiments}
\label{others}

\subsection{Experimental setup}

\paragraph{Datasets}
We conduct our experiments on four benchmark datasets with various sizes and document lengths, 
including \textit{20Newsgroups} \textbf{(20NG)} \cite{lang1995newsweeder}, \textit{Tag My News} \textbf{(TMN)} \cite{vitale2012classification}, 
\textit{WikiText-103} \textbf{(WIKI)} \cite{merity2016pointer}, and \textit{Reuters Corpus Volume II} \textbf{(RCV2)} \cite{lewis2004rcv1}. 
The statistics of these datasets are presented in Table~\ref{datasets}.
In particular, TMN is a short text corpus with an average of about 20 words per document;
20NG and TMN are the two corpora that are associated with document labels.
\vspace{-2mm}

\paragraph{Baseline methods}
As baselines, we choose several exemplary ones from the state-of-the-art topic models, including: 1) \textbf{LDA} \cite{blei2003latent}, one of the most widely used topic models; 2) \textbf{ProdLDA} \cite{srivastava2017autoencoding}, an NTM which replaces the mixture model in LDA with a product of
experts; 3) \textbf{ETM} \cite{dieng2020topic}, an NTM that marries conventional topic models with word embeddings; 4) \textbf{WHAI} \cite{zhang2018whai}, a hierarchical NTM which develops a deep Weibull variational encoder based on PGBN \cite{zhou2016augmentable}; 
5) \textbf{SawETM} \cite{duan2021sawtooth}, which proposes a Sawtooth Connection module to build the dependencies between topics at different layers; 6) \textbf{TopicNet} \cite{duan2021topicnet}, a knowledge-based hierarchical NTM that guides topic discovery through prior semantic graph.
All baselines are implemented meticulously according to their official code.

\begin{figure}[tb]
\centering
\setlength{\belowcaptionskip}{0.3cm}
\includegraphics[width=0.98\textwidth]{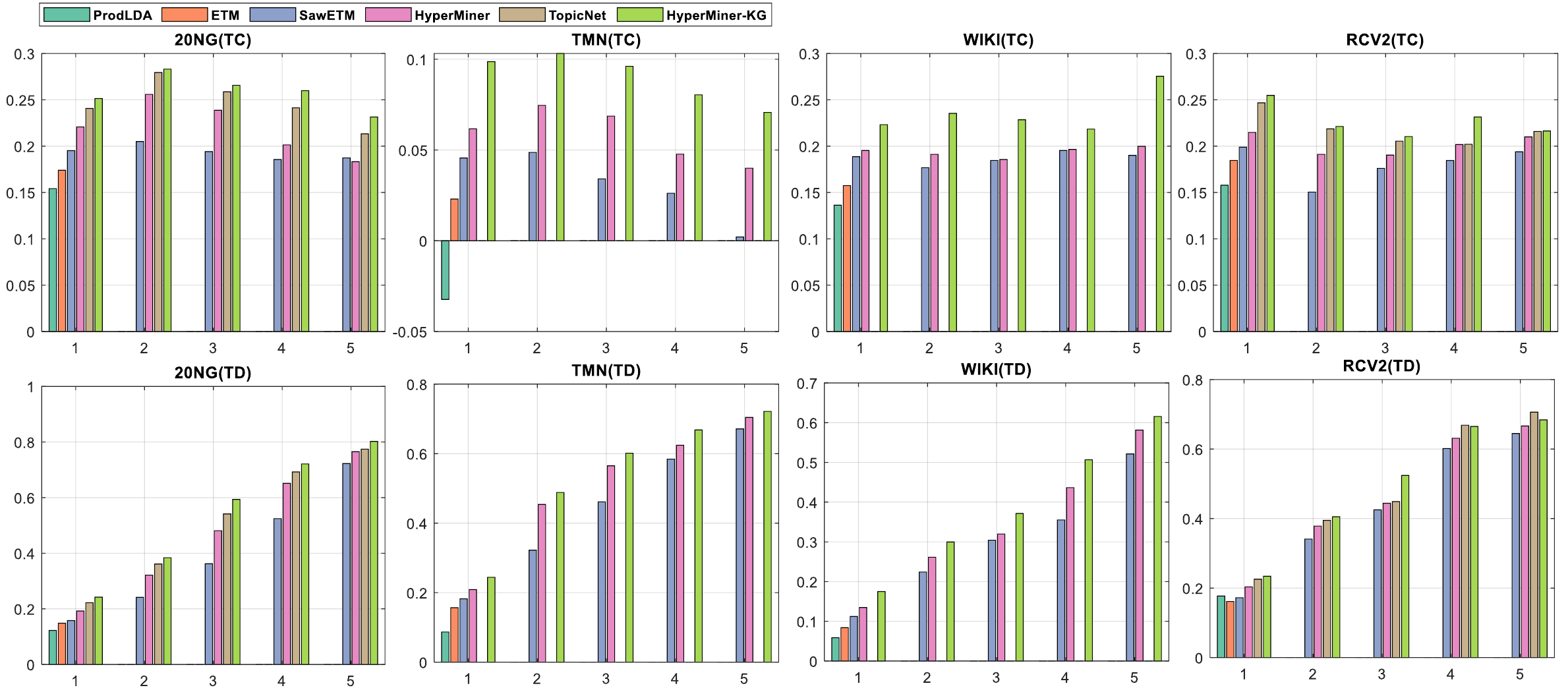}
\captionsetup{font={small}}
\caption{The performance comparison of different models on the topic quality. 
The top row shows the topic coherence score, $i.e.$, NPMI, and the bottom row displays the topic diversity score.
The horizontal axis represents the index of the layers and we set up 5 layers for all the hierarchical topic models.
The result of TopicNet on TMN and WIKI is missing because of memory overflow for large vocabulary size.}
\label{fig:topic_quality}
\end{figure}

\paragraph{Evaluation metrics}
We aim to evaluate our model's performance in terms of both topic quality and document representation. 
For topic quality, we adopt topic coherence (\textbf{TC}) and topic diversity (\textbf{TD}) as performance metrics. 
Given a reference corpus, TC measures the interpretability of each topic by computing the semantic coherence of the most significant words \cite{mimno2011optimizing}. 
Precisely, we apply the widely used Normalized Pointwise Mutual Information (NPMI) \cite{aletras2013evaluating} and compute it over the top 10 words of each topic, 
with the original document collections of each dataset serving as the reference corpus.
Note that the value of NPMI ranges from $-1$ to 1, and higher values indicate better interpretability.
TD, as the name suggests, measures how diverse the discovered topics are.
Following \citet{dieng2020topic}, we define TD to be the percentage of unique words in the top 25 words of all learned topics.
TD close to 0 means redundant topics and TD close to 1 implies more varied topics.

On the other hand, since per-document topic proportions can be viewed as unsupervised document representations,
we intend to evaluate the quality of such representations by performing document clustering tasks.
We report the purity and Normalized Mutual Information (NMI) \cite{manning2010introduction} on two datasets providing the document labels, $i.e.$, 20NG and TMN.
Concretely, with the default training/test split of each dataset, we first train a topic model on the training set, 
and then the trained model is used to extract features $\bm{\theta}$ of all test documents. 
Subsequently, we apply the KMeans algorithm on $\bm{\theta}$ and calculate the purity and NMI of the KMeans clusters (denoted by \textbf{km-Purity} and \textbf{km-NMI}). 
Note that both of the two metrics range from 0 to 1, and higher scores indicate better performance.
For the hierarchical topic models, we take latent units $\bm{\theta}^{(1)}$ of the first layer as the document feature. 

\subsection{Experimental results}  

\paragraph{Topic quality}
Considering that not all discovered topics are interpretable \cite{zhao2018dirichlet},
we select the top 50\% topics with the highest NPMI values and report the average score over those selected topics 
to evaluate the topic quality comprehensively. 
Figure~\ref{fig:topic_quality} exhibits the performance comparison results of different models.
Note that HyperMiner is the variant of SawETM that replaces the inner product between Euclidean embeddings with the distance between hyperbolic embeddings,
corresponding to the model  proposed in Section~\ref{sec31}.
While HyperMiner-KG is an advanced version of HyperMiner that guides the learning of topic taxonomy by external structural knowledge, 
as  introduced in Section~\ref{sec32}.
From what is shown in Figure~\ref{fig:topic_quality}, 
HyperMiner achieves consistent performance gains on all datasets compared to SawETM, in regard of both TC and TD,
which demonstrates the superiority of hyperbolic geometry in uncovering the latent hierarchies among topics and words.
In addition, as knowledge-guided topic models, both TopicNet and HyperMiner-KG get better performance than those without any supervision,
indicating the positive role of prior knowledge in helping to mine more interpretable and diverse topics.
However, HyperMiner-KG still performs slightly better than TopicNet while consuming less memory.
We attribute this result to our naturally-designed framework of injecting the tree-structure knowledge in a contrastive manner.

\begin{table}
  \caption{km-Purity and km-NMI for document clustering. 
  The best and second best scores of each dataset are highlighted in boldface and with an underline, respectively. 
  The embedding dimension for embedded topic models is set as 50.}
  \label{clustering}
  \centering
  \small
  \renewcommand{\arraystretch}{1.25} 
  \begin{tabular}{ p{2.5cm}<{\centering} p{2.3cm}<{\centering} p{2.3cm}<{\centering} p{2.3cm}<{\centering} p{2.3cm}<{\centering}}
  \toprule
   \multirow{2}*{Method}&
   \multicolumn{2}{c}{20NG}& \multicolumn{2}{c}{TMN}\cr
   \cmidrule(lr){2-3} \cmidrule(lr){4-5}
   &km-Purity&km-NMI&km-Purity&km-NMI\cr
   \midrule
   LDA \cite{blei2003latent}   &38.43 $\pm$ 0.52  &35.98 $\pm$ 0.39    &48.17 $\pm$ 0.69 &30.96 $\pm$ 0.78  \\
   ProdLDA \cite{srivastava2017autoencoding}   &39.21 $\pm$ 0.63    &36.52 $\pm$ 0.51    &55.28 $\pm$ 0.67   &35.57 $\pm$ 0.72  \\   
   ETM \cite{dieng2020topic}  &42.68 $\pm$ 0.71    &37.72 $\pm$ 0.64    &59.35 $\pm$ 0.74    &38.75 $\pm$ 0.86  \\
   WHAI \cite{zhang2018whai}  &40.89 $\pm$ 0.35 & 38.90 $\pm$ 0.27   &58.06 $\pm$ 0.45 &37.34 $\pm$ 0.48\\
   SawETM \cite{duan2021sawtooth}  &43.36 $\pm$ 0.48    &41.59 $\pm$ 0.62     &61.13 $\pm$ 0.56 &40.78 $\pm$ 0.63\\
   TopicNet \cite{duan2021topicnet}   &42.94 $\pm$ 0.41    &40.76 $\pm$  0.53    &60.52 $\pm$ 0.50   &40.09 $\pm$ 0.54\\
  \midrule
   HyperETM     &43.63 $\pm$ 0.51   &39.06 $\pm$ 0.64  &61.22 $\pm$ 0.62   & 40.52 $\pm$ 0.71 \\
   HyperMiner    &\underline{44.37} $\pm$ 0.38    &\underline{42.83} $\pm$ 0.45  &\underline{62.96} $\pm$ 0.48  &\underline{41.93} $\pm$ 0.52\\
   HyperMiner-KG   &\textbf{45.16} $\pm$ 0.35     &\textbf{43.65} $\pm$ 0.39    &\textbf{63.84} $\pm$ 0.43  &\textbf{42.81} $\pm$ 0.47 \\
  \bottomrule
 \end{tabular}
 \vspace{-0.2cm}
\end{table}

\begin{table}
  \caption{Accuracy for document classification on 20NG, with different embedding dimensions for embedded topic models.}
  \label{classification}
  \renewcommand{\arraystretch}{1.25}
  \centering
  \small
  \begin{tabular}{ p{2.2cm}<{\centering} p{1.8cm}<{\centering} p{1.8cm}<{\centering} p{1.8cm}<{\centering} p{1.8cm}<{\centering} p{1.8cm}<{\centering}}
    \toprule
    &\  $D=2$     &$D=5$    &$D=10$    &$D=20$    &$D=50$    \\
    \cmidrule(r){2-6}
    ETM \cite{dieng2020topic}      & 19.87 $\pm$ 0.81   & 33.64 $\pm$ 0.69    & 39.06 $\pm$ 0.54    & 42.13 $\pm$ 0.47     & 43.85  $\pm$ 0.42  \\
    HyperETM    & \textbf{24.33} $\pm$ 0.76   & \textbf{36.57} $\pm$ 0.65      &\textbf{40.92} $\pm$ 0.56      & \textbf{43.04}  $\pm$ 0.43   &\textbf{44.38} $\pm$ 0.40  \\
    \midrule
    SawETM \cite{duan2021sawtooth}       & 16.74 $\pm$ 0.78    & 27.05 $\pm$ 0.66   & 31.68 $\pm$ 0.51     & 34.06 $\pm$ 0.42   & 35.42 $\pm$ 0.37\\
    HyperMiner   & \textbf{20.16} $\pm$ 0.80  &  \textbf{29.73} $\pm$ 0.63  & \textbf{33.04} $\pm$ 0.49   & \textbf{34.98} $\pm$ 0.41     & \textbf{36.01} $\pm$ 0.36  \\
    \midrule
    TopicNet \cite{duan2021topicnet}   & 20.29 $\pm$ 0.58  &31.26 $\pm$ 0.51        & 34.57 $\pm$ 0.45  & 36.84 $\pm$ 0.39     & 38.02 $\pm$ 0.36   \\
    HyperMiner-KG   & \textbf{22.83} $\pm$ 0.55  & \textbf{33.15} $\pm$ 0.50       & \textbf{36.28} $\pm$ 0.43      & \textbf{38.11} $\pm$ 0.40    & \textbf{39.46} $\pm$ 0.34   \\
    \bottomrule
  \end{tabular}
  \vspace{0.3cm}
\end{table}

\paragraph{Document representation}
Table~\ref{clustering} shows the clustering performance of different models.
We run all the models in comparison five times with different random seeds and report the mean and standard deviation. 
From the results presented above, we have the following remarks: 
\textbf{i)} For all the evaluation metrics, our proposed improved variants perform consistently better than their prototypical 
models (refer to HyperETM versus ETM, and HyperMiner versus SawETM), which demonstrates that the introduced hyperbolic 
embeddings are beneficial to both the discovery of high-quality topics and the learning of good document representations. 
\textbf{ii)} As a knowledge-guided topic model, HyperMiner-KG achieves a significant improvement over the base model 
SawETM, while TopicNet suffers a slight performance degradation compared to SawETM, which also serves as its base 
model in the original paper. This observation shows that with the hyperbolic contrastive loss, 
our model not only injects the knowledge successfully into the learning of hierarchical topics, 
but also achieves a better balance among the comprehensive metrics of topic modeling.  
\textbf{iii)} The superior performance of our model on TMN also suggests its potential for short text topic modeling. 

To further investigate the effectiveness of our method under different dimensional settings, 
we proceed to compare the extrinsic predictive performance of document representations through classification tasks. 
Consistent with the practice in clustering tasks, 
we first collect the features of training set $\bm{\theta}_{tr}$ and test set $\bm{\theta}_{te}$ separately, 
which are inferred by a well-trained topic model. 
Then we train an SVM classifier using $\bm{\theta}_{tr}$ and their corresponding labels.
Finally, we use the trained classifier to predict the labels of $\bm{\theta}_{te}$ and compute the accuracy.
Table~\ref{classification} illustrates the classification results of different embedded topic models.
From the table we can see that the improved variants with our method surpass their base counterparts in various dimensionality settings. 
Especially, the performance gap between them has been further widened in the low-dimensional embedding space, 
confirming the natural advantage of hyperbolic distance metric in learning useful document representations.
\vspace{-0.2cm}

\begin{figure}[tb]
\centering
\includegraphics[width=\textwidth]{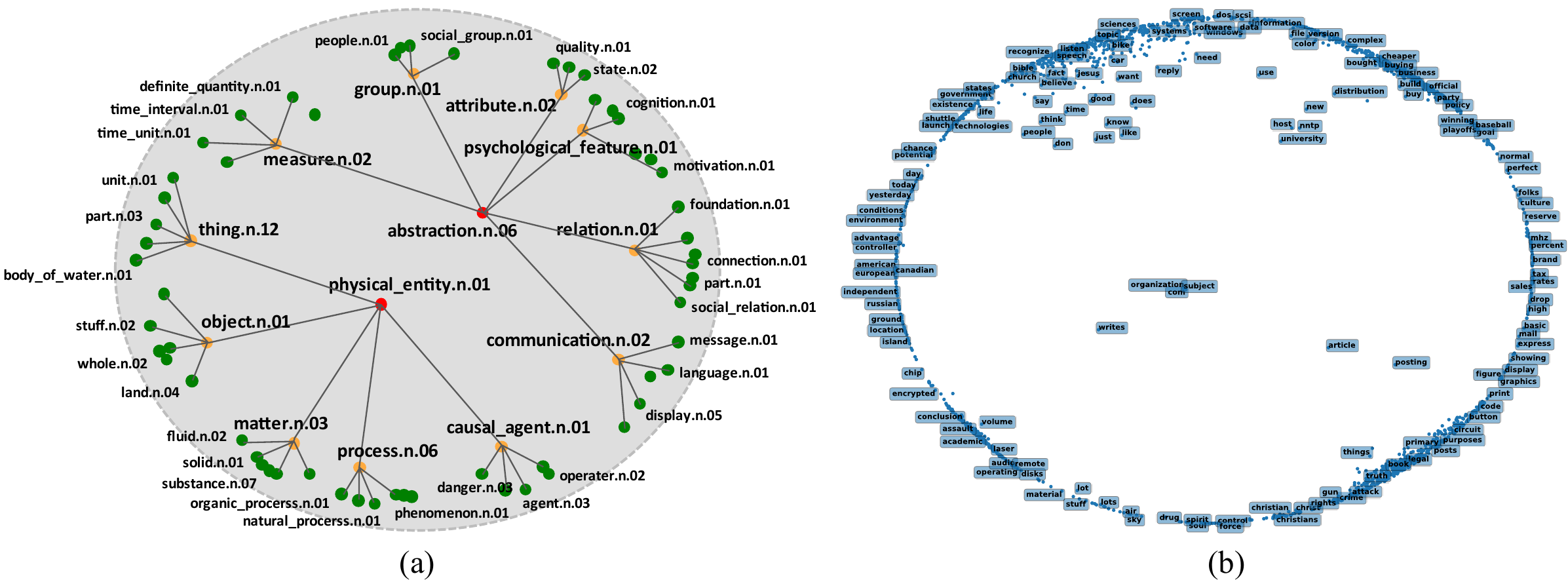}
\captionsetup{font={small}}
\caption{Visualization of 2D hyperbolic embedding space learned by HyperMiner-KG. 
(a) concept hierarchy: topic embeddings and their corresponding prior semantic concepts, where different coloured points represent topic embeddings at different layers.
(b) lexical hierarchy: word embeddings and their corresponding meanings.}
\label{fig:hyperbolic_space}
\end{figure}

\begin{figure}[h]
\centering
\setlength{\belowcaptionskip}{0.3cm}
\includegraphics[width=\textwidth]{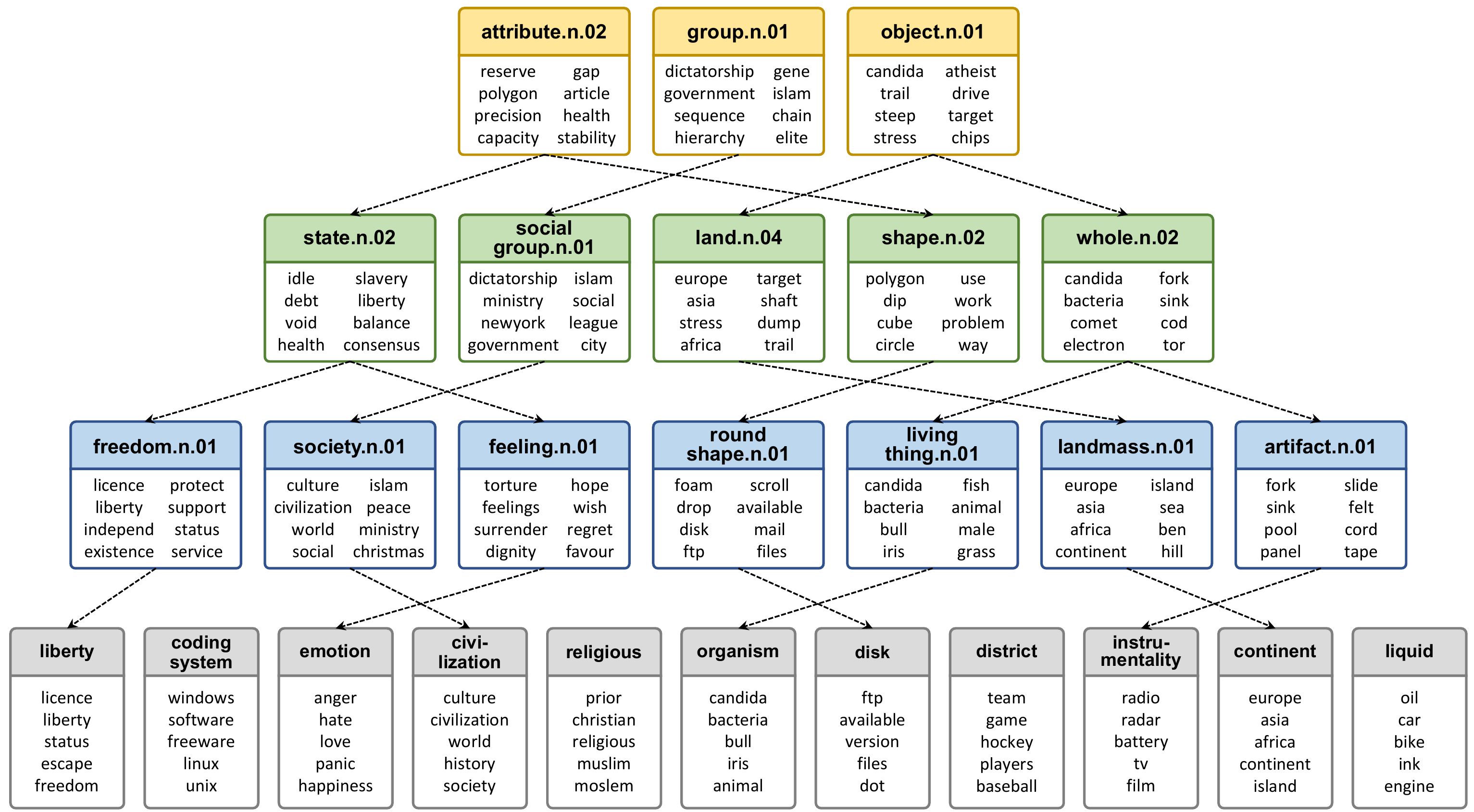}
\captionsetup{font={small}}
\caption{Illustration of the topic taxonomy learned by a 5-layer HyperMiner-KG on \textit{20Newsgroups}, we show the contents of some example topics, 
which are learned under the guidance of prior semantic concepts. Note that the example topics are selected from the bottom 4 layers,
with different colored boxes to distinguish them.}
\label{fig:topic_visualization}
\end{figure}

\paragraph{Visualization of embedding space and topics}
As our proposed HyperMiner-KG imposes a prior hierarchical constraint ($i.e.$, concept taxonomy) on the embedding space,
the topic embeddings and word embeddings are learned to maintain this structure as much as possible. 
Therefore, to verify the effectiveness of our proposed regularization term for injecting structural knowledge,
we visualize the two-dimensional hyperbolic embedding space learned by HyperMiner-KG, as displayed in Figure~\ref{fig:hyperbolic_space}.
Figure~\ref{fig:hyperbolic_space}(a) exhibits the learned topic embeddings and the concept hierarchy used to guide them, 
in which we can see that the distribution of topic embeddings well preserves the semantic structure of prior knowledge.
Specifically, for topics guided by higher-level and more general concepts ($e.g.$, \textit{physical\_entity.n.01}), 
their embeddings tend to locate in the center of the disc. While for those led by slightly more certain concepts ($e.g.$, \textit{substance.n.07}),
their embeddings prefer to scatter around the boundary.
Figure~\ref{fig:hyperbolic_space}(b) presents the distribution of learned word embeddings, 
which also reflects the underlying lexical hierarchy of the corpus.
The words that co-occur more frequently with different 
terms ($e.g.$, \textit{organization}, \textit{subject}) tend to fall around the center of the disc, 
so as to maintain a small distance from arbitrary words. 
In contrast, those words with precise meanings fall near the edge area with spacious room 
and keep a small distance only from words with similar semantics.

Furthermore, to qualitatively demonstrate the crucial role of prior structural knowledge in helping to discover more interpretable topics,
we show the contents ($i.e.$, top words) of some learned topics by HyperMiner-KG, as illustrated in Figure~\ref{fig:topic_visualization}.
From it we can observe that in a majority of cases the prior concepts can successfully guide the topics to learn semantically consistent contents
($e.g.$, the topics guided by \textit{coding\_system} and \textit{instrumentality}). 
Moreover, the contents of topics at different layers are also semantically related due to the concepts guiding them. For instance, the content of the topic guided by \textit{whole.n.02} covers the contents of topics led by \textit{living\_thing.n.01} and \textit{artifact.n.01}, respectively.
Another interesting phenomenon is that the topic led by \textit{round\_shape.n.01} involves not only words related to shapes, but also some other words such as \textit{files} and \textit{ftp}. The reason could be one of its child concepts \textit{disk} often co-occurs with those words in the given corpus, suggesting the topic learning is co-influenced by both data likelihood and knowledge regularization.

\section{Related Work}
Historically, many attempts have been made to develop hierarchical topic models.
Expanding on their flat counterparts, 
hierarchical topic models aim to generate an intuitive topic hierarchy by capturing the correlations among topics. 
Due to the inflexibility of requiring accurate posterior inference, 
early works \cite{griffiths2003hierarchical, blei2006correlated, paisley2014nested, zhou2016augmentable} 
primarily focused on learning potential topic hierarchies purely from data, 
with some additional probabilistic assumptions being imposed.
Also, there is a small body of work that tries to integrate domain knowledge into the process of discovering topic hierarchies.
For example, anchored CorEx \cite{gallagher2017anchored} takes user-provided seed words and 
learns informative topics by maximizing total correlation while preserving anchor words related information.
More recently, JoSH \cite{meng2020hierarchical} adopts a more effective strategy
that takes a category hierarchy as the guidance and models category-word semantic correlation via joint spherical text and tree embedding.
Different from anchored CorEx and JoSH, which deviate from the conventional topic modeling framework,
our approach still follows the regular probabilistic generative process.
In addition, we use a concept taxonomy to guide the topic learning 
so that more fine-grained topics can be mined. 
In this regard, our proposed HyperMiner-KG is much more related to TopicNet \cite{duan2021topicnet}, 
yet it is more efficient with a smaller storage footprint.

\section{Conclusion}
This paper presents a novel framework that introduces hyperbolic embeddings to represent words and topics on top of existing embedded topic models. 
By using the hyperbolic distance to measure the semantic similarity between words and topics, 
the model can better explore the underlying semantic hierarchy to find more interpretable topics. 
Besides, a hyperbolic contrastive loss has been further proposed, 
which effectively injects prior structural knowledge into hierarchical topic models to guide learning a meaningful topic taxonomy. 
Our method shows appealing properties that can overcome several shortcomings of existing embedded topic models. 
Extensive experiments have been carried out, demonstrating that our method achieves consistent performance improvements 
in discovering high-quality topics and deriving useful document representations.

\begin{ack}

Bo Chen acknowledges the support of NSFC (U21B2006 and 61771361), 
Shaanxi Youth Innovation Team Project, the 111 Project (No. B18039) 
and the Program for Oversea Talent by Chinese Central Government.
\end{ack}

\bibliography{neurips_2022}

\begin{thebibliography}{60}
\providecommand{\natexlab}[1]{#1}
\providecommand{\url}[1]{\texttt{#1}}
\expandafter\ifx\csname urlstyle\endcsname\relax
  \providecommand{\doi}[1]{doi: #1}\else
  \providecommand{\doi}{doi: \begingroup \urlstyle{rm}\Url}\fi

\bibitem[Wang et~al.(2007)Wang, McCallum, and Wei]{wang2007topical}
Xuerui Wang, Andrew McCallum, and Xing Wei.
\newblock Topical n-grams: Phrase and topic discovery, with an application to
  information retrieval.
\newblock In \emph{Seventh IEEE international conference on data mining (ICDM
  2007)}, pages 697--702. IEEE, 2007.

\bibitem[Mimno et~al.(2009)Mimno, Wallach, Naradowsky, Smith, and
  McCallum]{mimno2009polylingual}
David Mimno, Hanna Wallach, Jason Naradowsky, David~A Smith, and Andrew
  McCallum.
\newblock Polylingual topic models.
\newblock In \emph{Proceedings of the 2009 conference on empirical methods in
  natural language processing}, pages 880--889, 2009.

\bibitem[Rubin et~al.(2012)Rubin, Chambers, Smyth, and
  Steyvers]{rubin2012statistical}
Timothy~N Rubin, America Chambers, Padhraic Smyth, and Mark Steyvers.
\newblock Statistical topic models for multi-label document classification.
\newblock \emph{Machine learning}, 88\penalty0 (1):\penalty0 157--208, 2012.

\bibitem[Jiang et~al.(2015)Jiang, Qian, Shen, Fu, and Mei]{jiang2015author}
Shuhui Jiang, Xueming Qian, Jialie Shen, Yun Fu, and Tao Mei.
\newblock Author topic model-based collaborative filtering for personalized
  {POI} recommendations.
\newblock \emph{IEEE transactions on multimedia}, 17\penalty0 (6):\penalty0
  907--918, 2015.

\bibitem[Wang et~al.(2018)Wang, Gan, Wang, Shen, Huang, Ping, Satheesh, and
  Carin]{wang2018topic}
Wenlin Wang, Zhe Gan, Wenqi Wang, Dinghan Shen, Jiaji Huang, Wei Ping, Sanjeev
  Satheesh, and Lawrence Carin.
\newblock Topic compositional neural language model.
\newblock In \emph{International Conference on Artificial Intelligence and
  Statistics}, pages 356--365. PMLR, 2018.

\bibitem[Jelodar et~al.(2020)Jelodar, Wang, Orji, and Huang]{jelodar2020deep}
Hamed Jelodar, Yongli Wang, Rita Orji, and Shucheng Huang.
\newblock Deep sentiment classification and topic discovery on novel
  coronavirus or covid-19 online discussions: Nlp using lstm recurrent neural
  network approach.
\newblock \emph{IEEE Journal of Biomedical and Health Informatics}, 24\penalty0
  (10):\penalty0 2733--2742, 2020.

\bibitem[Blei et~al.(2003)Blei, Ng, and Jordan]{blei2003latent}
David~M Blei, Andrew~Y Ng, and Michael~I Jordan.
\newblock Latent {D}irichlet allocation.
\newblock \emph{Journal of machine Learning research}, 3\penalty0
  (Jan):\penalty0 993--1022, 2003.

\bibitem[Griffiths et~al.(2003)Griffiths, Jordan, Tenenbaum, and
  Blei]{griffiths2003hierarchical}
Thomas Griffiths, Michael Jordan, Joshua Tenenbaum, and David Blei.
\newblock Hierarchical topic models and the nested {C}hinese restaurant
  process.
\newblock \emph{Advances in neural information processing systems}, 16, 2003.

\bibitem[Blei and Lafferty(2006)]{blei2006correlated}
David Blei and John Lafferty.
\newblock Correlated topic models.
\newblock \emph{Advances in neural information processing systems},
  18:\penalty0 147, 2006.

\bibitem[Mcauliffe and Blei(2007)]{mcauliffe2007supervised}
Jon Mcauliffe and David Blei.
\newblock Supervised topic models.
\newblock \emph{Advances in neural information processing systems}, 20, 2007.

\bibitem[Zhou et~al.(2012)Zhou, Hannah, Dunson, and Carin]{zhou2012beta}
Mingyuan Zhou, Lauren Hannah, David Dunson, and Lawrence Carin.
\newblock Beta-negative binomial process and {P}oisson factor analysis.
\newblock In \emph{Artificial Intelligence and Statistics}, pages 1462--1471.
  PMLR, 2012.

\bibitem[Paisley et~al.(2014)Paisley, Wang, Blei, and
  Jordan]{paisley2014nested}
John Paisley, Chong Wang, David~M Blei, and Michael~I Jordan.
\newblock Nested hierarchical {D}irichlet processes.
\newblock \emph{IEEE transactions on pattern analysis and machine
  intelligence}, 37\penalty0 (2):\penalty0 256--270, 2014.

\bibitem[Kingma and Welling(2013)]{kingma2013auto}
Diederik~P Kingma and Max Welling.
\newblock Auto-encoding variational {B}ayes.
\newblock \emph{arXiv preprint arXiv:1312.6114}, 2013.

\bibitem[Rezende et~al.(2014)Rezende, Mohamed, and
  Wierstra]{rezende2014stochastic}
Danilo~Jimenez Rezende, Shakir Mohamed, and Daan Wierstra.
\newblock Stochastic backpropagation and approximate inference in deep
  generative models.
\newblock In \emph{International conference on machine learning}, pages
  1278--1286. PMLR, 2014.

\bibitem[Miao et~al.(2016)Miao, Yu, and Blunsom]{miao2016neural}
Yishu Miao, Lei Yu, and Phil Blunsom.
\newblock Neural variational inference for text processing.
\newblock In \emph{International conference on machine learning}, pages
  1727--1736. PMLR, 2016.

\bibitem[Srivastava and Sutton(2017)]{srivastava2017autoencoding}
Akash Srivastava and Charles Sutton.
\newblock Autoencoding variational inference for topic models.
\newblock \emph{arXiv preprint arXiv:1703.01488}, 2017.

\bibitem[Zhang et~al.(2018)Zhang, Chen, Guo, and Zhou]{zhang2018whai}
Hao Zhang, Bo~Chen, Dandan Guo, and Mingyuan Zhou.
\newblock Whai: Weibull hybrid autoencoding inference for deep topic modeling.
\newblock \emph{arXiv preprint arXiv:1803.01328}, 2018.

\bibitem[Nan et~al.(2019)Nan, Ding, Nallapati, and Xiang]{nan2019topic}
Feng Nan, Ran Ding, Ramesh Nallapati, and Bing Xiang.
\newblock Topic modeling with {W}asserstein autoencoders.
\newblock \emph{arXiv preprint arXiv:1907.12374}, 2019.

\bibitem[Mikolov et~al.(2013{\natexlab{a}})Mikolov, Chen, Corrado, and
  Dean]{mikolov2013efficient}
Tomas Mikolov, Kai Chen, Greg Corrado, and Jeffrey Dean.
\newblock Efficient estimation of word representations in vector space.
\newblock \emph{arXiv preprint arXiv:1301.3781}, 2013{\natexlab{a}}.

\bibitem[Mikolov et~al.(2013{\natexlab{b}})Mikolov, Sutskever, Chen, Corrado,
  and Dean]{mikolov2013distributed}
Tomas Mikolov, Ilya Sutskever, Kai Chen, Greg~S Corrado, and Jeff Dean.
\newblock Distributed representations of words and phrases and their
  compositionality.
\newblock \emph{Advances in neural information processing systems}, 26,
  2013{\natexlab{b}}.

\bibitem[Petterson et~al.(2010)Petterson, Buntine, Narayanamurthy, Caetano, and
  Smola]{petterson2010word}
James Petterson, Wray Buntine, Shravan Narayanamurthy, Tib{\'e}rio Caetano, and
  Alex Smola.
\newblock Word features for latent {D}irichlet allocation.
\newblock \emph{Advances in Neural Information Processing Systems}, 23, 2010.

\bibitem[Nguyen et~al.(2015)Nguyen, Billingsley, Du, and
  Johnson]{nguyen2015improving}
Dat~Quoc Nguyen, Richard Billingsley, Lan Du, and Mark Johnson.
\newblock Improving topic models with latent feature word representations.
\newblock \emph{Transactions of the Association for Computational Linguistics},
  3:\penalty0 299--313, 2015.

\bibitem[Li et~al.(2016)Li, Wang, Zhang, Sun, and Ma]{li2016topic}
Chenliang Li, Haoran Wang, Zhiqian Zhang, Aixin Sun, and Zongyang Ma.
\newblock Topic modeling for short texts with auxiliary word embeddings.
\newblock In \emph{Proceedings of the 39th International ACM SIGIR conference
  on Research and Development in Information Retrieval}, pages 165--174, 2016.

\bibitem[Zhao et~al.(2018{\natexlab{a}})Zhao, Du, Buntine, and
  Zhou]{zhao2018inter}
He~Zhao, Lan Du, Wray Buntine, and Mingyuan Zhou.
\newblock Inter and intra topic structure learning with word embeddings.
\newblock In \emph{International Conference on Machine Learning}, pages
  5892--5901. PMLR, 2018{\natexlab{a}}.

\bibitem[Wang et~al.()Wang, Guo, Zhao, Zheng, Tanwisuth, Chen, and Zhou]{wete}
Dongsheng Wang, Dandan Guo, He~Zhao, Huangjie Zheng, Korawat Tanwisuth,
  Bo~Chen, and Mingyuan Zhou.
\newblock Representing mixtures of word embeddings with mixtures of topic
  embeddings.
\newblock In \emph{The Tenth International Conference on Learning
  Representations, {ICLR} 2022, Virtual Event, April 25-29, 2022}.

\bibitem[Bianchi et~al.(2020)Bianchi, Terragni, and Hovy]{bianchi2020pre}
Federico Bianchi, Silvia Terragni, and Dirk Hovy.
\newblock Pre-training is a hot topic: Contextualized document embeddings
  improve topic coherence.
\newblock \emph{arXiv preprint arXiv:2004.03974}, 2020.

\bibitem[Dieng et~al.(2020)Dieng, Ruiz, and Blei]{dieng2020topic}
Adji~B Dieng, Francisco~JR Ruiz, and David~M Blei.
\newblock Topic modeling in embedding spaces.
\newblock \emph{Transactions of the Association for Computational Linguistics},
  8:\penalty0 439--453, 2020.

\bibitem[Duan et~al.(2021{\natexlab{a}})Duan, Wang, Chen, Wang, Chen, Li, Ren,
  and Zhou]{duan2021sawtooth}
Zhibin Duan, Dongsheng Wang, Bo~Chen, Chaojie Wang, Wenchao Chen, Yewen Li, Jie
  Ren, and Mingyuan Zhou.
\newblock Sawtooth factorial topic embeddings guided gamma belief network.
\newblock In \emph{International Conference on Machine Learning}, pages
  2903--2913. PMLR, 2021{\natexlab{a}}.

\bibitem[Nickel and Kiela(2017)]{nickel2017poincare}
Maximillian Nickel and Douwe Kiela.
\newblock Poincar\'e embeddings for learning hierarchical representations.
\newblock \emph{Advances in neural information processing systems}, 30, 2017.

\bibitem[Nickel et~al.(2014)Nickel, Jiang, and Tresp]{nickel2014reducing}
Maximilian Nickel, Xueyan Jiang, and Volker Tresp.
\newblock Reducing the rank in relational factorization models by including
  observable patterns.
\newblock \emph{Advances in Neural Information Processing Systems}, 27, 2014.

\bibitem[Duan et~al.(2021{\natexlab{b}})Duan, Xu, Chen, Wang, Zhou,
  et~al.]{duan2021topicnet}
Zhibin Duan, Yishi Xu, Bo~Chen, Chaojie Wang, Mingyuan Zhou, et~al.
\newblock Topicnet: Semantic graph-guided topic discovery.
\newblock \emph{Advances in Neural Information Processing Systems}, 34,
  2021{\natexlab{b}}.

\bibitem[Gromov(1987)]{gromov1987hyperbolic}
Mikhael Gromov.
\newblock Hyperbolic groups.
\newblock In \emph{Essays in group theory}, pages 75--263. Springer, 1987.

\bibitem[Hamann(2018)]{hamann2018tree}
Matthias Hamann.
\newblock On the tree-likeness of hyperbolic spaces.
\newblock In \emph{Mathematical proceedings of the cambridge philosophical
  society}, volume 164, pages 345--361. Cambridge University Press, 2018.

\bibitem[Ganea et~al.(2018)Ganea, B{\'e}cigneul, and
  Hofmann]{ganea2018hyperbolic}
Octavian Ganea, Gary B{\'e}cigneul, and Thomas Hofmann.
\newblock Hyperbolic entailment cones for learning hierarchical embeddings.
\newblock In \emph{International Conference on Machine Learning}, pages
  1646--1655. PMLR, 2018.

\bibitem[Sala et~al.(2018)Sala, De~Sa, Gu, and R{\'e}]{sala2018representation}
Frederic Sala, Chris De~Sa, Albert Gu, and Christopher R{\'e}.
\newblock Representation tradeoffs for hyperbolic embeddings.
\newblock In \emph{International conference on machine learning}, pages
  4460--4469. PMLR, 2018.

\bibitem[Tifrea et~al.(2018)Tifrea, B{\'e}cigneul, and
  Ganea]{tifrea2018poincar}
Alexandru Tifrea, Gary B{\'e}cigneul, and Octavian-Eugen Ganea.
\newblock Poincar\'e glove: Hyperbolic word embeddings.
\newblock \emph{arXiv preprint arXiv:1810.06546}, 2018.

\bibitem[Cho et~al.(2019)Cho, DeMeo, Peng, and Berger]{cho2019large}
Hyunghoon Cho, Benjamin DeMeo, Jian Peng, and Bonnie Berger.
\newblock Large-margin classification in hyperbolic space.
\newblock In \emph{The 22nd international conference on artificial intelligence
  and statistics}, pages 1832--1840. PMLR, 2019.

\bibitem[Atchison and Shen(1980)]{atchison1980logistic}
J~Atchison and Sheng~M Shen.
\newblock Logistic-normal distributions: Some properties and uses.
\newblock \emph{Biometrika}, 67\penalty0 (2):\penalty0 261--272, 1980.

\bibitem[Lee(2013)]{lee2013smooth}
John~M Lee.
\newblock Smooth manifolds.
\newblock In \emph{Introduction to Smooth Manifolds}, pages 1--31. Springer,
  2013.

\bibitem[Nickel and Kiela(2018)]{nickel2018learning}
Maximillian Nickel and Douwe Kiela.
\newblock Learning continuous hierarchies in the lorentz model of hyperbolic
  geometry.
\newblock In \emph{International Conference on Machine Learning}, pages
  3779--3788. PMLR, 2018.

\bibitem[Zhou et~al.(2016)Zhou, Cong, and Chen]{zhou2016augmentable}
Mingyuan Zhou, Yulai Cong, and Bo~Chen.
\newblock Augmentable gamma belief networks.
\newblock \emph{The Journal of Machine Learning Research}, 17\penalty0
  (1):\penalty0 5656--5699, 2016.

\bibitem[Wang et~al.(2022)Wang, Xu, Li, Duan, Wang, Chen, and
  Zhou]{wang2022knowledge}
Dongsheng Wang, Yishi Xu, Miaoge Li, Zhibin Duan, Chaojie Wang, Bo~Chen, and
  Mingyuan Zhou.
\newblock Knowledge-aware {B}ayesian deep topic model.
\newblock \emph{arXiv preprint arXiv:2209.14228}, 2022.

\bibitem[Miller(1995)]{miller1995wordnet}
George~A Miller.
\newblock Wordnet: A lexical database for english.
\newblock \emph{Communications of the ACM}, 38\penalty0 (11):\penalty0 39--41,
  1995.

\bibitem[You et~al.(2020)You, Chen, Sui, Chen, Wang, and Shen]{you2020graph}
Yuning You, Tianlong Chen, Yongduo Sui, Ting Chen, Zhangyang Wang, and Yang
  Shen.
\newblock Graph contrastive learning with augmentations.
\newblock \emph{Advances in Neural Information Processing Systems},
  33:\penalty0 5812--5823, 2020.

\bibitem[Hassani and Khasahmadi(2020)]{hassani2020contrastive}
Kaveh Hassani and Amir~Hosein Khasahmadi.
\newblock Contrastive multi-view representation learning on graphs.
\newblock In \emph{International Conference on Machine Learning}, pages
  4116--4126. PMLR, 2020.

\bibitem[Zhu et~al.(2020)Zhu, Xu, Yu, Liu, Wu, and Wang]{zhu2020deep}
Yanqiao Zhu, Yichen Xu, Feng Yu, Qiang Liu, Shu Wu, and Liang Wang.
\newblock Deep graph contrastive representation learning.
\newblock \emph{arXiv preprint arXiv:2006.04131}, 2020.

\bibitem[Peng et~al.(2020)Peng, Huang, Luo, Zheng, Rong, Xu, and
  Huang]{peng2020graph}
Zhen Peng, Wenbing Huang, Minnan Luo, Qinghua Zheng, Yu~Rong, Tingyang Xu, and
  Junzhou Huang.
\newblock Graph representation learning via graphical mutual information
  maximization.
\newblock In \emph{Proceedings of The Web Conference 2020}, pages 259--270,
  2020.

\bibitem[Lang(1995)]{lang1995newsweeder}
Ken Lang.
\newblock Newsweeder: Learning to filter netnews.
\newblock In \emph{Machine Learning Proceedings 1995}, pages 331--339.
  Elsevier, 1995.

\bibitem[Vitale et~al.(2012)Vitale, Ferragina, and
  Scaiella]{vitale2012classification}
Daniele Vitale, Paolo Ferragina, and Ugo Scaiella.
\newblock Classification of short texts by deploying topical annotations.
\newblock In \emph{European Conference on Information Retrieval}, pages
  376--387. Springer, 2012.

\bibitem[Merity et~al.(2016)Merity, Xiong, Bradbury, and
  Socher]{merity2016pointer}
Stephen Merity, Caiming Xiong, James Bradbury, and Richard Socher.
\newblock Pointer sentinel mixture models.
\newblock \emph{arXiv preprint arXiv:1609.07843}, 2016.

\bibitem[Lewis et~al.(2004)Lewis, Yang, Russell-Rose, and Li]{lewis2004rcv1}
David~D Lewis, Yiming Yang, Tony Russell-Rose, and Fan Li.
\newblock {RCV1}: A new benchmark collection for text categorization research.
\newblock \emph{Journal of machine learning research}, 5\penalty0
  (Apr):\penalty0 361--397, 2004.

\bibitem[Mimno et~al.(2011)Mimno, Wallach, Talley, Leenders, and
  McCallum]{mimno2011optimizing}
David Mimno, Hanna Wallach, Edmund Talley, Miriam Leenders, and Andrew
  McCallum.
\newblock Optimizing semantic coherence in topic models.
\newblock In \emph{Proceedings of the 2011 conference on empirical methods in
  natural language processing}, pages 262--272, 2011.

\bibitem[Aletras and Stevenson(2013)]{aletras2013evaluating}
Nikolaos Aletras and Mark Stevenson.
\newblock Evaluating topic coherence using distributional semantics.
\newblock In \emph{Proceedings of the 10th International Conference on
  Computational Semantics (IWCS 2013)--Long Papers}, pages 13--22, 2013.

\bibitem[Manning et~al.(2010)Manning, Raghavan, and
  Sch{\"u}tze]{manning2010introduction}
Christopher Manning, Prabhakar Raghavan, and Hinrich Sch{\"u}tze.
\newblock Introduction to information retrieval.
\newblock \emph{Natural Language Engineering}, 16\penalty0 (1):\penalty0
  100--103, 2010.

\bibitem[Zhao et~al.(2018{\natexlab{b}})Zhao, Du, Buntine, and
  Zhou]{zhao2018dirichlet}
He~Zhao, Lan Du, Wray Buntine, and Mingyuan Zhou.
\newblock Dirichlet belief networks for topic structure learning.
\newblock \emph{Advances in neural information processing systems}, 31,
  2018{\natexlab{b}}.

\bibitem[Gallagher et~al.(2017)Gallagher, Reing, Kale, and
  Ver~Steeg]{gallagher2017anchored}
Ryan~J Gallagher, Kyle Reing, David Kale, and Greg Ver~Steeg.
\newblock Anchored correlation explanation: Topic modeling with minimal domain
  knowledge.
\newblock \emph{Transactions of the Association for Computational Linguistics},
  5:\penalty0 529--542, 2017.

\bibitem[Meng et~al.(2020)Meng, Zhang, Huang, Zhang, Zhang, and
  Han]{meng2020hierarchical}
Yu~Meng, Yunyi Zhang, Jiaxin Huang, Yu~Zhang, Chao Zhang, and Jiawei Han.
\newblock Hierarchical topic mining via joint spherical tree and text
  embedding.
\newblock In \emph{Proceedings of the 26th ACM SIGKDD international conference
  on knowledge discovery \& data mining}, pages 1908--1917, 2020.

\bibitem[S{\o}nderby et~al.(2016)S{\o}nderby, Raiko, Maal{\o}e, S{\o}nderby,
  and Winther]{sonderby2016ladder}
Casper~Kaae S{\o}nderby, Tapani Raiko, Lars Maal{\o}e, S{\o}ren~Kaae
  S{\o}nderby, and Ole Winther.
\newblock Ladder variational autoencoders.
\newblock \emph{Advances in neural information processing systems}, 29, 2016.

\bibitem[Kingma and Ba(2014)]{kingma2014adam}
Diederik~P Kingma and Jimmy Ba.
\newblock Adam: A method for stochastic optimization.
\newblock \emph{arXiv preprint arXiv:1412.6980}, 2014.

\bibitem[Ungar(2007)]{ungar2007hyperbolic}
Abraham~A Ungar.
\newblock The hyperbolic square and mobius transformations.
\newblock \emph{Banach Journal of Mathematical Analysis}, 1\penalty0
  (1):\penalty0 101--116, 2007.

\end{thebibliography}
\bibliographystyle{unsrtnat}
\newpage

\clearpage
\section*{Checklist}


\begin{enumerate}

\item For all authors...
\begin{enumerate}
  \item Do the main claims made in the abstract and introduction accurately reflect the paper's contributions and scope?
    \answerYes{}
  \item Did you describe the limitations of your work?
    \answerYes{We describe the limitations of our word in the appendix}
  \item Did you discuss any potential negative societal impacts of your work?
    \answerYes{We discuss the negative societal impacts in the appendix}
  \item Have you read the ethics review guidelines and ensured that your paper conforms to them?
    \answerYes{}
\end{enumerate}

\item If you are including theoretical results...
\begin{enumerate}
  \item Did you state the full set of assumptions of all theoretical results?
    \answerYes{}
        \item Did you include complete proofs of all theoretical results?
    \answerNA{}
\end{enumerate}

\item If you ran experiments...
\begin{enumerate}
  \item Did you include the code, data, and instructions needed to reproduce the main experimental results (either in the supplemental material or as a URL)?
    \answerYes{} They will be included in the supplementary material.
  \item Did you specify all the training details ($e.g.$, data splits, hyperparameters, how they were chosen)?
    \answerYes{}
        \item Did you report error bars ($e.g.$, with respect to the random seed after running experiments multiple times)?
    \answerYes{}
        \item Did you include the total amount of compute and the type of resources used ($e.g.$, type of GPUs, internal cluster, or cloud provider)?
    \answerYes{}
\end{enumerate}

\item If you are using existing assets ($e.g.$, code, data, models) or curating/releasing new assets...
\begin{enumerate}
  \item If your work uses existing assets, did you cite the creators?
    \answerYes{}
  \item Did you mention the license of the assets?
    \answerNA{}
  \item Did you include any new assets either in the supplemental material or as a URL?
    \answerNo{}
  \item Did you discuss whether and how consent was obtained from people whose data you're using/curating?
    \answerNA{}
  \item Did you discuss whether the data you are using/curating contains personally identifiable information or offensive content?
    \answerNA{}
\end{enumerate}

\item If you used crowdsourcing or conducted research with human subjects...
\begin{enumerate}
  \item Did you include the full text of instructions given to participants and screenshots, if applicable?
    \answerNA{}
  \item Did you describe any potential participant risks, with links to Institutional Review Board (IRB) approvals, if applicable?
    \answerNA{}
  \item Did you include the estimated hourly wage paid to participants and the total amount spent on participant compensation?
    \answerNA{}
\end{enumerate}

\end{enumerate}


\appendix
\newpage

\section{Discussions}

\subsection{Limitations}
We in this paper propose a method to improve existing embedded topic models (ETMs) by introducing 
the hyperbolic distance to measure the semantic similarity between topics and words. 
Additionally, to mine a meaningful topic taxonomy with the guidance of prior structural knowledge, 
we further develop a regularization term based on contrastive learning
that can effectively inject prior knowledge into hierarchical topic models. 
The main limitation of our work could be the mismatch problem between the given prior knowledge and the target corpus. 
Specifically, to provide proper guidance for mining an interpretable topic taxonomy, 
the prior structural knowledge should be well matched with the corresponding dataset.
Although we present a seemingly effective heuristic strategy by finding ancestor concepts of each word in the vocabulary,
there are certainly better ways to construct qualified priors to guide the learning of topic hierarchies. 
However, this is beyond the scope of this paper and we will conduct a thorough investigation of this issue in future work.

\subsection{Broader impact}
Our work builds on advanced topic modeling techniques and thus can be used for regular text analysis.
For example, topic discovery and obtaining document representation.
Furthermore, our work also provides a solution to inject prior knowledge as an inductive bias to
influence topic learning, which is particularly useful when users are only 
interested in certain types of information. 
Imagine a user's goal is to extract the parts about a specific topic from a large amount of news, 
our model can act as a good filter.
Or consider the application scenario of recommending papers to researchers, 
the browsing history as prior knowledge reflects their preferences,
which can be incorporated into the model so that only the papers on related topics are  presented, 
thus improving the recommendation accuracy.
Potential negative societal impact of our work could arise from malicious intent in 
changing model's behavior by injecting deliberate human prejudice, which
may harm the fairness of the community. 
However, we hope our work is utilized to enable new downstream applications primarily
from the originality of benefiting the community development.

\section{Implementations}\label{appendB}

\renewcommand\thefigure{\Alph{section}. \arabic{figure}}    
\setcounter{figure}{0}
\renewcommand\thetable{\Alph{section}. \arabic{table}}    
\setcounter{table}{0}

\subsection{Data  splits}
We summarize the training/test split of each dataset in Table~\ref{datasets_split}. 
In particular, 20NG\footnote{\url{http://qwone.com/~jason/20Newsgroups/}} and 
TMN\footnote{\url{http://acube.di.unipi.it/tmn-dataset/}} are used to evaluate both topic quality and document representation,
and their document collections are divided into standard training sets and test sets.
WIKI\footnote{\url{https://blog.salesforceairesearch.com/the-wikitext-long-term-dependency-language-modeling-dataset/}} and RCV2\footnote{\url{https://trec.nist.gov/data/reuters/reuters.html}} are only used for topic discovery, 
so we use all documents for training.
\begin{table}[!h]
  \caption{Splits of the datasets}
  \label{datasets_split}
  \setlength{\tabcolsep}{8pt} 
  \renewcommand{\arraystretch}{1.25} 
  \centering
  \small
  \begin{tabular}{cccc}
    \toprule
    &Vocabulary size     &Number of training docs      &Number of test docs \\
    \cmidrule(r){2-4}
    20NG      &8,581         &11,314         &7,532   \\
    TMN        &13,368      &26,077          &6,520 \\
    WIKI        &20,000      &28,472        & /  \\
    RCV2       &7,282        &804,414      & / \\
    \bottomrule
  \end{tabular}
\end{table}

\subsection{Prior concept taxonomy}\label{B2}
\begin{figure*}[!htb]
\begin{center}
\includegraphics[width=0.68\textwidth]{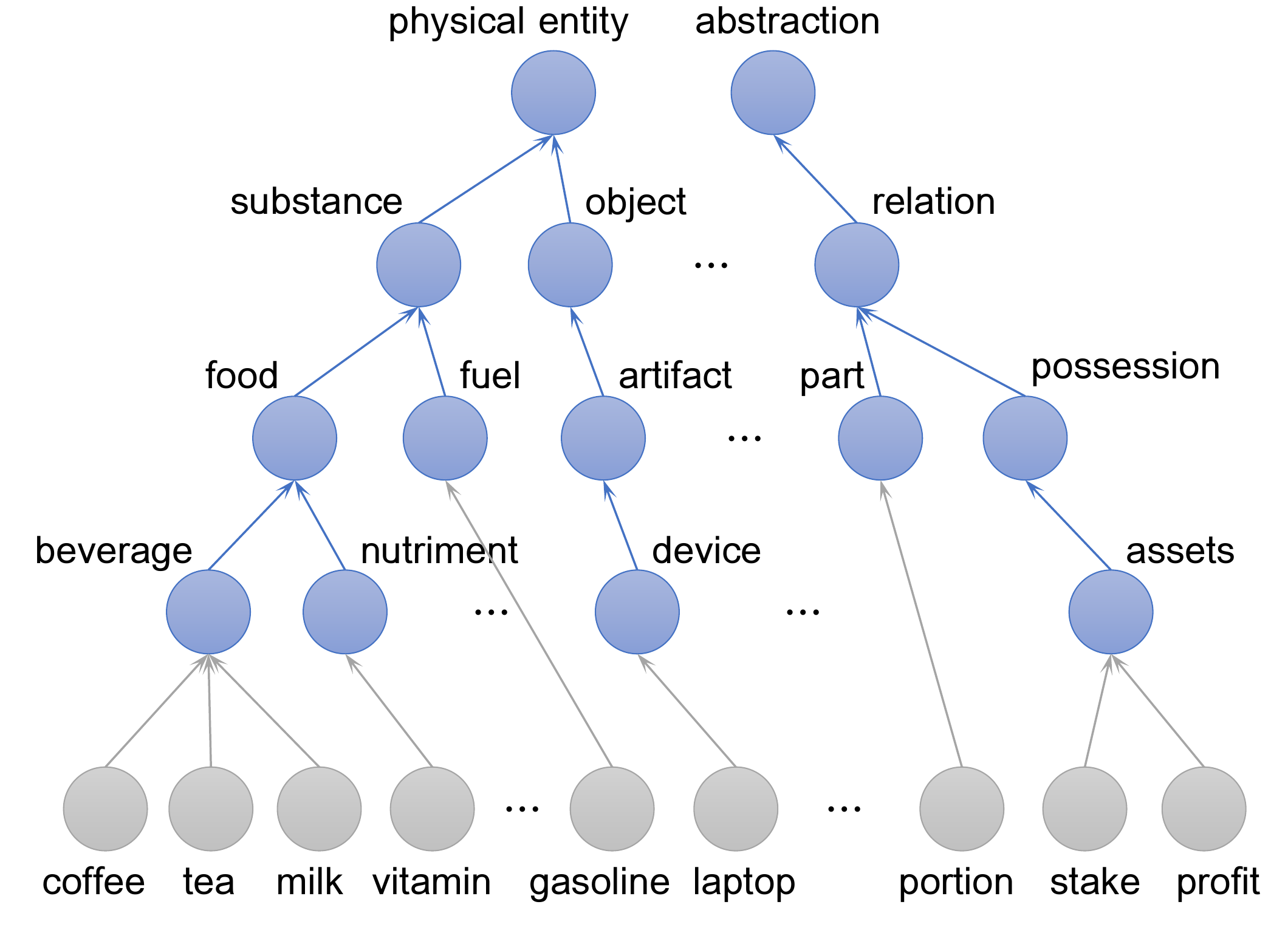}
\captionsetup{font={small}}
\caption{Construction of a concept taxonomy. 
Gray circles represent leaf nodes corresponding to the words in the vocabulary, 
blue circles are the ancestor nodes on the words' hypernym paths, 
each of which is defined by a semantic concept.
}
\label{fig:concept_tree}
\end{center}
\end{figure*}

In this part, we give an example to illustrate how the prior concept taxonomy (or structural knowledge) is constructed.
Specifically, given the vocabulary of a dataset, we first filter out those words that are not included in the WordNet thesaurus.
For each of the remaining words, we then find its ancestor concepts along the hypernym paths integrated in WordNet
($e.g.$, for the word \textit{coffee}, it has a hypernym path where a series of abstract concepts, $i.e.$, \textit{beverage}, \textit{food}, \textit{substance}, \textit{physical\_entity}, successively appear, as displayed in Figure~\ref{fig:concept_tree}).
After traversing all the words, we can get a concept tree with great depth, but the number of nodes in the deepest layer may be very small.
Therefore, to keep the number of nodes growing as the layer gets deeper, we choose to reserve the sevaral layers closest to the root node.
For the words in the deeper layers, we connect them directly to their ancestor concepts of the deepest layer that has been preserved.

\vspace{-2mm}
In this way, we construct a concept taxonomy with a depth of 5 (the layer of words is excluded) for each dataset.
More precisely, the number of concepts in each layer is [2, 12, 83, 325, 560] for 20NG; 
[2, 11, 84, 366, 683] for TMN; [2, 12, 91, 408, 810] for WIKI; [2, 11, 70, 306, 540] for RCV2.

\subsection{Inference network}
\begin{figure*}[!htb]
\begin{center}
\includegraphics[width=0.8\textwidth]{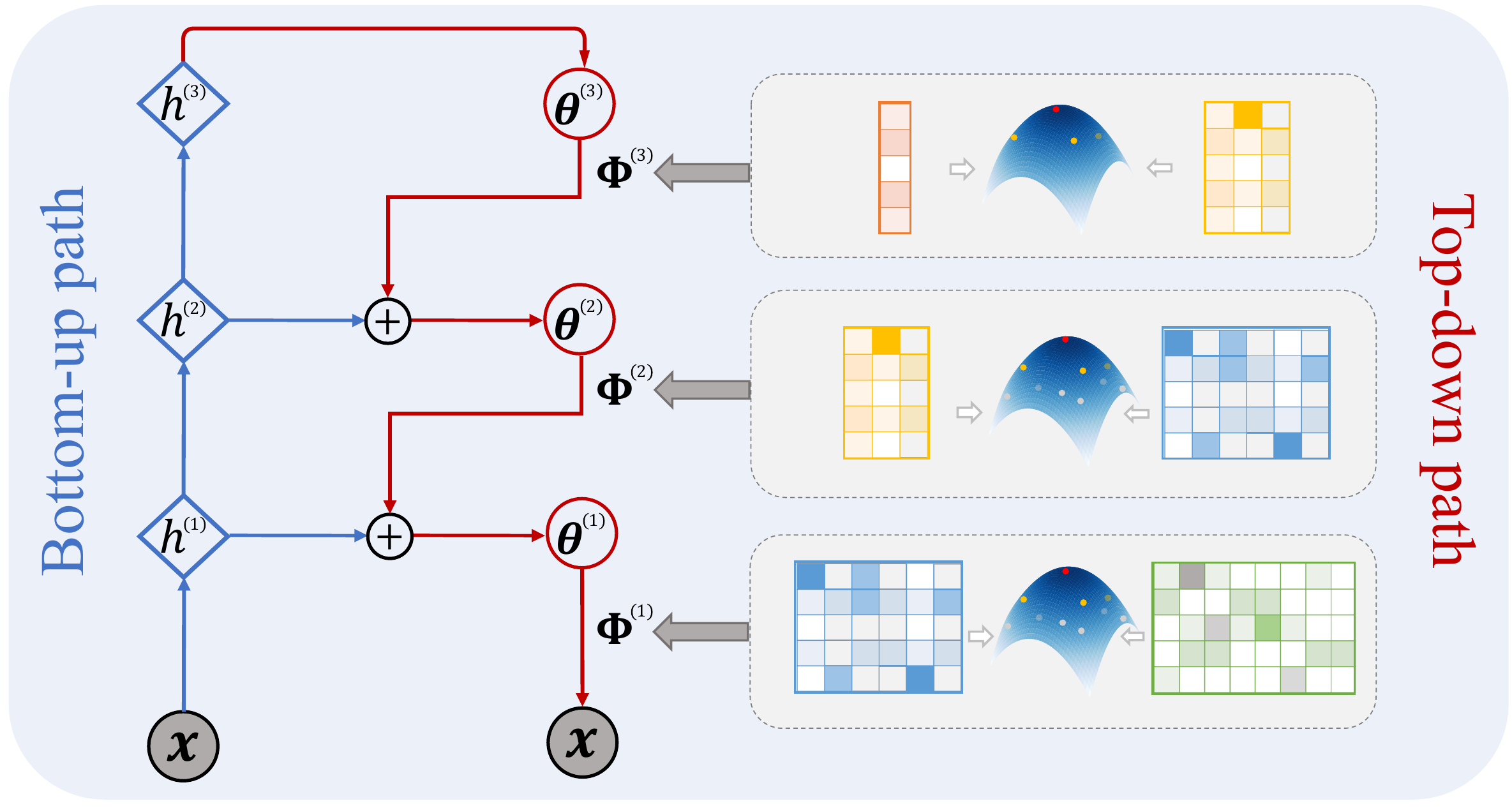}
\captionsetup{font={small}}
\caption{Overview of the inference network. 
The bottom-up path propagates the likelihood information from the original input,
and the top-down path conducts the prior information from the generative model.}
\label{fig:inference_net}
\end{center}
\end{figure*}

Since the exact posterior distribution for $\bm{\theta}^{(l)}$ is intractable in our generative model,
we aim to design a sampling-based inference network to approximate the true posterior distribution, 
which is adopted by most neural topic models.
In view of the hierarchical structure where deep-layer latent variables are difficult to receive effective information from the original input,
we draw experience from LadderVAE \cite{sonderby2016ladder} and use a skip-connected deterministic upward path 
to infer the hidden features of the input $\bm{x}$
\begin{equation} \label{deterministic}\small
\begin{split}
& \bm{h}_j^{\left( 1 \right)} = \mathrm{MLP} \left( \bm{x}\right), \\
& \bm{h}_j^{\left( l \right)} = \bm{h}_j^{\left( l-1 \right)} + \mathrm{MLP} \left( \bm{h}_j^{\left( l-1 \right)} \right), \\
\end{split}
\end{equation}
where MLP is a multi-layer perceptron consisting of two fully connected layers,
with the ReLU activation following behind. 
The obtained hidden features are subsequently combined with the prior from the stochastic up-down
path to approximate the variational posterior, which is expressed as
\begin{equation} \label{inference}\small
\begin{split}
& \bm{k}_j^{\left( l \right)} = \mathrm{Softplus} \left( \mathrm{Linear} \left( \bm{\Phi}^{(l+1)}\bm{\theta}_j^{(l+1)} \oplus \bm{h}_j^{(l)} \right)\right), \\
& \bm{t}_j^{\left( l \right)} = \mathrm{Softplus} \left( \mathrm{Linear} \left( \bm{\Phi}^{(l+1)}\bm{\theta}_j^{(l+1)} \oplus \bm{h}_j^{(l)} \right)\right), \\
& q\left( \bm{\theta}_j^{(l)} \vert \bm{h}_j^{(l)}, \bm{\theta}_j^{(l+1)}, \bm{\Phi}^{l+1}\right) = \mathrm{Weibull}\left( \bm{k}_j^{(l)}, \bm{t}_j^{(l)} \right), \\
\end{split}
\end{equation}
where $\oplus$ denotes the concatenation at topic dimension, 
$\mathrm{Linear}$ is a simple fully connected layer with identity activation, 
and $\mathrm{Softplus}$ applies $\mathrm{log}(1 + \mathrm{exp}(\cdotv))$ nonlinearity to each element, 
ensuring that shape and scale parameters of the Weibull distribution are positive.
The reason for using the Weibull distribution to approximate the gamma-distributed conditional posterior 
has been explained in the main body. 
Note that both shape and scale parameters, $i.e.$, $\bm{k}_j^{(l)} \in \mathbb{R}^{K_l}$ and 
$\bm{t}_j^{(l)} \in \mathbb{R}^{K_l}$, are inferred through the neural networks, 
by using the combination of the bottom-up likelihood information and the top-down prior information as input. 
Figure~\ref{fig:inference_net} depicts the overall inference process.

\subsection{Training protocal}
All our experiments are performed on a single Nvidia Geforce RTX 3090 GPU card,
with PyTorch as the programming platform to implement our models. 
For the MLP module in the inference network,
we set the number of hidden neurons as 300.
In addition, we also add a batch normalization layer to prevent overfitting.
For all the embedded topic models, we set the embedding size as 50.
To optimize our models, we use the Adam \cite{kingma2014adam} optimizer with a learning rate of 0.01. 
As for the size of each  mini- batch, we set it to 200 for all datasets.
What's more, for our proposed HyperMiner-KG, the size of negative samples is set as 256
for each anchor to calculate the hyperbolic contrastive loss.

It is also worth noting that the number of topics at each layer in hierarchical topic models
is set to be consistent with the number of concepts at the corresponding layer in the constructed concept taxonomy.
Please refer to Section~\ref{B2} for detailed settings. For the single-layer topic models, we set the number of topics 
to be the same as the number of concepts at the deepest layer of the concept taxonomy.

\section{Hyperbolic Space}\label{appendC}

\subsection{Equivalence between Poincaré Ball model and Lorentz model}
\begin{figure}[tb]
\centering
\includegraphics[width=0.8\textwidth]{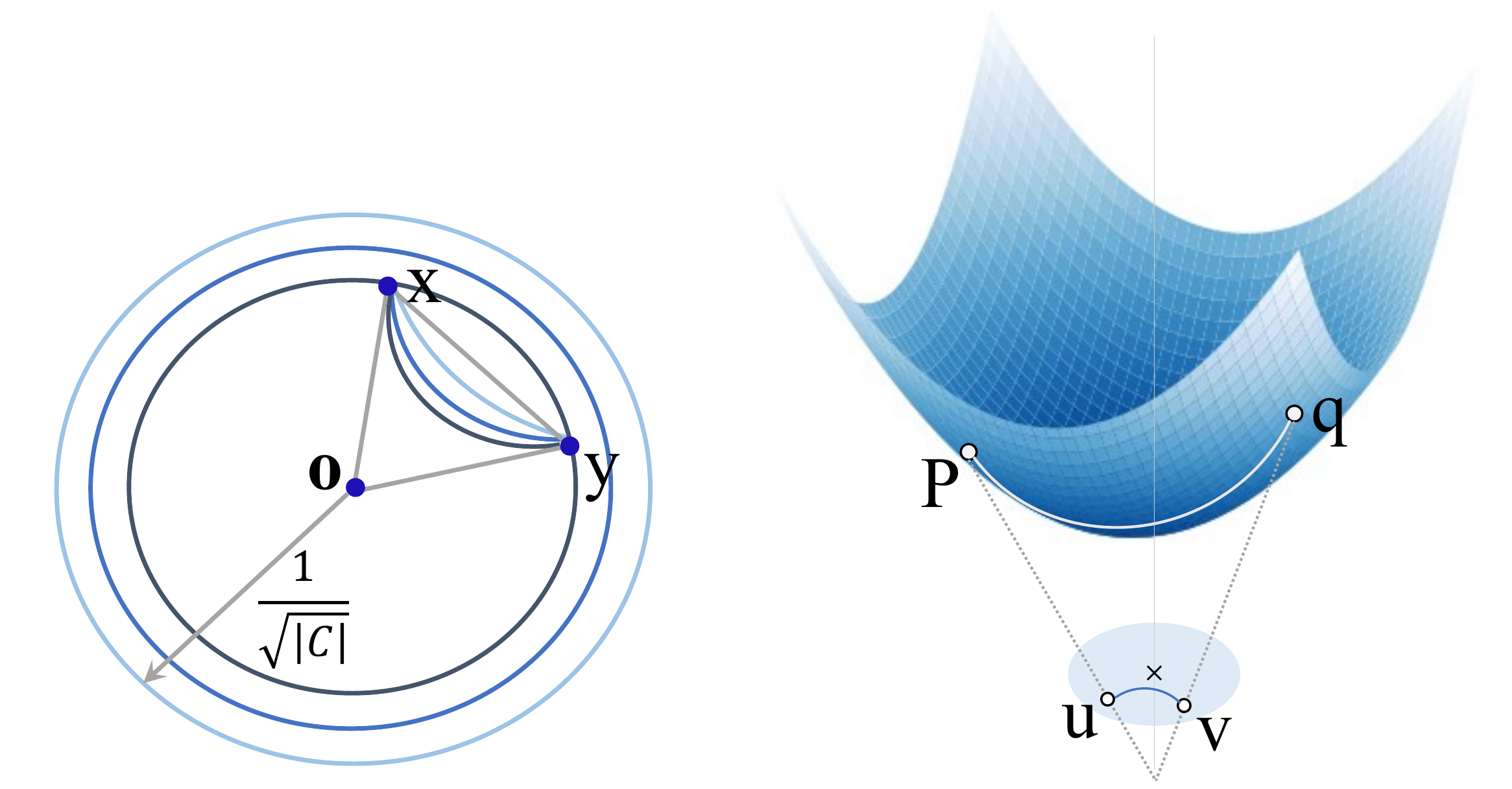}
\captionsetup{font={small}}
\caption{Left: Poincaré disk geodesics (shortest path) connecting x and y for different curvatures. As curvature decreases, the distance between x and y increases, and the geodesics lines get closer to the origin. Right: points (p, q) lie on the surface of a two-dimensional Lorentz space, points (u, v) are the mapping of (p, q) onto the two-dimensinal Poincaré disk. Note that (p, q) are points in three-dimensional space.}
\label{poincare_vs_lorentz}
\end{figure}

In this section, we aim to offer an intuitive explanation about the equivalence of the two hyperbolic models mentioned in the main text.
Firstly, we need to clarify the concept of geodesic. In geometry, a geodesic is commonly a curve representing the shortest path between two points in a surface. In a typical Euclidean space, the geodesic is the straight line connecting two points, and its length is the widely used Euclidean distance 
that is determined only by the coordinates of the two points. While in a hyperbolic space, the length of a geodesic is not only related to the coordinates of its connected points, but also affected by the curvature of hyperbolic space. This can be illustrated by the left side of Figure~\ref{poincare_vs_lorentz},
as the curvature (negative) decreases, the corresponding curvature radius decreases, but the distance between x and y increases and 
the geodesics lines get closer to the origin.

The right side of Figure~\ref{poincare_vs_lorentz} clearly describes the projection of the geodesic on the Lorentz surface to the geodesic in the Poincaré disk.
We say that the two models are mathematically equivalent because points on the Poincare disk and points in the Lorentz space can be mapped to each other,
while all geometric properties including isometry are preserved. For example, to map a point in the Lorentz model into the corresponding point in the Poincaré ball, we have the following diffeomorphism \cite{nickel2018learning} $p: \mathcal{H}^n \rightarrow \mathcal{P}^n$, where
\begin{equation}
    p(x_0, x_1, \cdots, x_n) = \frac{(x_1, \cdots, x_n)}{x_0 + 1}
\end{equation}
Furthermore, points in $\mathcal{P}^n$ can be mapped back into $\mathcal{H}^n$ via
\begin{equation}
    p^{-1}(x_1, \cdots, x_n) = \frac{(1 + {\Vert x \Vert}^2, 2{x_1}, \cdots, 2{x_n})}{1 - {\Vert x \Vert}^2}
\end{equation}
To calculate the lengths of geodesics in the Poincaré disk and Lorentz space, respectively, please refer to the definition of $d_\mathcal{P}(\cdotv)$ and $d_\mathcal{L}(\cdotv)$ in Eq~(\ref{Hyp-Dist}). However, despite the mathematical equivalence of the two models, it does not mean that the lengths calculated by $d_\mathcal{P}(\cdotv)$ and $d_\mathcal{L}(\cdotv)$ are exactly the same.

\subsection{Related operations}
A Riemannian manifold $\left( \mathcal{M}, g \right)$ is a differentiable manifold $\mathcal{M}$ equipped with a metric tensor $g$.
It can be locally approximated to a linear Euclidean space at an arbitrary point $\bm{x} \in \mathcal{M}$, and the approximated space is termed
as a tangent space $\mathcal{T}_{\bm{x}} {\mathcal{M}}$. Hyperbolic spaces are smooth Riemannian manifolds with a constant negative curvature. 
There are several essential vector operations required for learning embeddings in a hyperbolic space, we will give an introduction to them in the following.

\textbf{Exponential and logarithmic maps}. 
An exponential map ${{\rm exp}_{\rm \bf x}}(\rm \bf v)$ is the function projecting a tangent vector
${\rm \bf v} \in {\mathcal{T}_{\bm{x}} \mathcal{M}} $ onto $\mathcal{M}$. 
A logarithmic map projects vectors on the manifold back to the tangent space satisfying
${\rm log}_{\rm \bf x}({\rm exp}_{\rm \bf x}(\rm \bf v)) = \rm \bf v$. 

\textbf{Parallel transport}. A parallel transport can move a tangent vector along the surface of a curved manifold.
For example, to move a tangent vector ${\rm \bf v} \in {\mathcal{T}_{\bm{x}}\mathcal{M}}$ to another tangent space $\mathcal{T}_{\bm{y}} \mathcal{M}$, we use the notation ${\rm PT}_{\rm \bf x \to \rm \bf y}^\mathcal{M}(\rm \bf v)$.

The concrete formula of these operations in Poincaré Ball and Lorentz model are summarized in Table~\ref{operations_summary}.
Where $\oplus$ and $\rm gyr[:; :]$ are the Möbius addition \cite{ungar2007hyperbolic} and gyration operator \cite{ungar2007hyperbolic}, respectively.

\begin{table}[ht]
  \caption{Summary of operations in the Poincaré ball model and the Lorentz model ($C = -1$)}
  \label{operations_summary}
\footnotesize
  \renewcommand{\arraystretch}{2} 
  \centering
  \begin{tabular}{p{1.2cm}<{\centering} p{5.8cm}<{\centering} p{5.6cm}<{\centering}}
  \toprule
   &\textbf{Poincaré Ball Model}   &\textbf{Lorentz Model} \cr
  \hline\hline
  \textbf{Log map}      &$\log_{\rm \bf x}^{\mathcal{P}}(\rm \bf y) = \frac{2}{\lambda_{\rm \bf x}} \rm artanh \left( \Vert {\rm \bf -x} \oplus \rm \bf y \Vert \right)\frac{\rm \bf -x \oplus \rm \bf y}{\Vert \rm \bf -x \oplus \rm \bf y \Vert}$             & $\log_{\rm \bf x}^{\mathcal{L}}(\rm \bf y) = \frac{\rm arcosh(-{\langle \rm \bf x, \rm \bf y \rangle}_\mathcal{L})}{\sqrt{{\langle \rm \bf x, \rm \bf y \rangle}_\mathcal{L}^{2} - 1}} (\rm \bf y + {\langle \rm \bf x, \rm \bf y \rangle}_\mathcal{L}\rm \bf x)$ \\
  \textbf{Exp map}      &$\exp_{\rm \bf x}^{\mathcal{P}}(\rm \bf v)=\rm \bf x \oplus \left( \rm tanh \left( \frac{ \lambda_{\rm \bf x}\Vert \rm \bf v \Vert}{2}\right) \frac{\rm \bf v}{\Vert \rm \bf v \Vert}  \right)$            &$\exp_{\rm \bf x}^{\mathcal{L}}(\rm \bf v)=\rm cosh ( {\Vert \rm \bf v \Vert}_\mathcal{L} ) \rm \bf x + \rm \bf v \frac{\rm sinh( {\Vert \rm \bf v \Vert}_\mathcal{L})}{{\Vert \rm \bf v \Vert}_\mathcal{L}}$\\
  \textbf{Transport}     &  ${\rm PT}_{\rm \bf x \to \rm \bf y}^\mathcal{P}(\rm \bf v) = \frac{\lambda_{\rm \bf x}}{\lambda_{\rm \bf y}} \rm gyr 
  [\rm \bf -x, \rm \bf y] \rm \bf v $      &  ${\rm PT}_{\rm \bf x \to \rm \bf y}^\mathcal{L}(\rm \bf v) = \rm \bf v + \frac{{\langle \rm \bf y, \rm \bf v\rangle}_\mathcal{L}}{1-{\langle \rm \bf x, \rm \bf y \rangle}_\mathcal{L}}(\rm \bf x + \rm \bf y) $\\ 
  \bottomrule
 \end{tabular}
\end{table}

\begin{table}
  \caption{km-Purity and km-NMI for document clustering, with different 
regularization coefficient for HyperMiner-KG. The best score of each
dataset is highlighted in boldface.}
  \label{clustering_ap}
  \small
  \renewcommand{\arraystretch}{1.25} 
  \centering
  \begin{tabular}{ p{2.5cm}<{\centering} p{2.3cm}<{\centering} p{2.3cm}<{\centering} p{2.3cm}<{\centering} p{2.3cm}<{\centering}}
  \toprule
   \multirow{2}*{HyperMiner-KG}&
   \multicolumn{2}{c}{20NG}& \multicolumn{2}{c}{TMN}\cr
   \cmidrule(lr){2-3} \cmidrule(lr){4-5}
   &km-Purity&km-NMI&km-Purity&km-NMI\cr
   \midrule
   $\lambda=0.1$ &43.76 $\pm$ 0.32  &42.63$\pm$ 0.38    &62.25 $\pm$ 0.46  &41.32 $\pm$ 0.49 \\
   $\lambda=1$   &44.13 $\pm$ 0.33    &42.96 $\pm$ 0.36    &62.73 $\pm$ 0.47   &41.68 $\pm$ 0.48 \\   
     $\lambda=2$ &44.48 $\pm$ 0.39    &43.28 $\pm$ 0.41    &63.07 $\pm$ 0.52    &42.06 $\pm$ 0.54  \\
    $\lambda=5$  &\textbf{45.16} $\pm$ 0.35 &\textbf{43.65} $\pm$ 0.39   &\textbf{63.84} $\pm$ 0.48 &\textbf{42.81} $\pm$ 0.52\\
   $\lambda=10$  &44.81 $\pm$ 0.37    &43.47 $\pm$ 0.38     &63.39 $\pm$ 0.46 &42.34 $\pm$ 0.50\\
  \bottomrule
 \end{tabular}
 \vspace{0.5cm}
\end{table}

\section{Additional Results}\label{appendD}

\subsection{Effect of regularization term}
To investigate the effect of the regularization term (prior structural knowledge) in HyperMiner-KG, we 
further evaluate the quality of document representations learned by HyperMiner-KG with different 
regularization coefficient $\lambda$ on document clustering tasks.

From the Table~\ref{clustering_ap} we can see, wth the increase of the regularization coefficient, 
HyperMiner-KG has shown better performance on both km-Purity and km-NMI, 
proving that incorporating prior structural knowledge is beneficial to learning better document representations.
However, the regularization coefficient is not the bigger the better, it has a most suitable value, which in our experiments is 5.

\end{document}